\documentclass[pra,preprint,showpacs,letterpaper]{revtex4-1}
\usepackage[english]{babel}
\usepackage{amsmath,amssymb,bm,psfrag,ifthen}
\newcommand{\vect}[1]{\bm{#1}}

\newcommand{\norm}[1]{|#1|}
\newcommand{\fou}[1]{\hat{#1}}
\newcommand{\feyn}[1]{\mathcal{D}[#1]}
\newcommand{\psic}{{\psi^\star}}
\newcommand{\psics}[1]{\psi^\star_{#1}}

\newcommand{\psia}{\psi}
\newcommand{\psias}[1]{\psia^{\phantom{\star}}_{#1}}

\newcommand{\ia}{j}
\newcommand{\ib}{k}
\newcommand*{\itemp}{\beta}
\newcommand*{\Q}{Q}
\newcommand*{\half}{\frac{1}{2}}

\newcommand*{\spin}{\sigma}
\newcommand*{\smp}{\epsilon}
\newcommand*{\ord}[1]{\mathcal{O}(#1)}
\newcommand*{\cd}{\chi_d}

\newcommand{\srv}[1]{\bar{#1}}
\newcommand*{\V}{V}
\newcommand*{\C}{C}
\newcommand*{\W}{W}
\newcommand*{\w}{w}

\newcommand*{\cp}{\mu}
\newcommand*{{\cps}}{\cp_\spin}

\newcommand*{\bbR}{\mathbb{R}}
\newcommand*{\bbZ}{\mathbb{Z}}
\newcommand*{\EQ}{E}

\newcommand*{\EH}{E_\textsc{h}}

\newcommand*{\EHX}{E_\textsc{hx}}
\newcommand*{\ETF}{E_\textsc{tf}}
\newcommand*{\EXC}{E_\textsc{x}}

\newcommand*{\ECR}{E_\textsc{c}}

\newcommand*{\ECA}{E_\textsc{c;1}}
\newcommand*{\ECS}{E_\textsc{c;1}^\star}
\newcommand*{\ECB}{E_\textsc{c;2}}

\newcommand*{\ECP}{\Delta}
\newcommand*{\fcp}{^\cp}

\newcommand*{\sEQ}{\srv{E}}
\newcommand*{\sECR}{\srv{E}_\textsc{c}}
\newcommand*{\sECRa}{\srv{E}_{\textsc{c};1}}
\newcommand*{\sECRb}{\srv{E}_{\textsc{c};2}}
\newcommand*{\sETF}{\srv{E}_\textsc{tf}}

\newcommand*{\sEHF}{\srv{E}_\textsc{hf}}

\newcommand*{\sEHXsm}{\srv{E}_{\textsc{hx};\text{s}}}

\newcommand*{\sEHXos}{\srv{E}_{\textsc{hx};\text{osc}}}

\newcommand*{\sEHX}{\srv{E}_\textsc{hx}}

\newcommand*{\GSL}{\lim_{\itemp\to\infty}}

\newcommand*{\PP}{P}
\newcommand*{\PSC}{P_\textsc{h}}
\newcommand*{\PXC}{P_\textsc{x}}

\newcommand*{\PCA}{P_\textsc{c;1}}
\newcommand*{\PCB}{P_\textsc{c;2}}

\newcommand*{\K}{K}

\newcommand*{\dos}{d}
\newcommand*{\dosos}{\dos_\text{osc}}
\newcommand*{\dossm}{\dos_\text{s}}
\newcommand*{\idos}{D}
\newcommand*{\den}{\rho}
\newcommand*{\deno}{\den_0}
\newcommand*{\nfo}{\nf_a}
\newcommand*{\nfi}{\nf_b}
\newcommand*{\Eoo}{E_{\textsc{h};a}}

\newcommand*{\Eoi}{E_{\textsc{h};b}}
\newcommand*{\cpo}{\cp}

\newcommand*{\cpoo}{\cp_a}
\newcommand*{\cpoop}{\cp_{a;+}}
\newcommand*{\cpoi}{\cp_b}

\newcommand*{\Wo}{\W_a}
\newcommand*{\Wi}{\W_b}
\newcommand*{\denoo}{\den_{0;a}}
\newcommand*{\denoi}{\den_{0;b}}
\newcommand*{\nf}{n}

\newcommand*{\kv}{\vect{k}}

\newcommand*{\pv}{\vect{p}}
\newcommand*{\qv}{\vect{q}}
\newcommand*{\rv}{\vect{r}}
\newcommand*{\sv}{\vect{s}}
\newcommand*{\tv}{\vect{t}}

\newcommand*{\xv}{\vect{x}}
\newcommand*{\yv}{\vect{y}}
\newcommand*{\zv}{\vect{z}}

\newcommand*{\heaviside}{\vartheta}
\newcommand*{\smhs}[1]{\heaviside(#1)}
\newcommand*{\hs}[1]{\heaviside\big(#1\big)}
\newcommand*{\kron}{\delta}
\newcommand*{\smks}[1]{\kron(#1)}
\newcommand*{\ks}[1]{\kron\big(#1\big)}

\newcommand*{\inv}{^{-1}}
\newcommand*{\FX}{(\xv)}
\newcommand*{\FY}{(\yv)}
\newcommand*{\FF}[2]{(#1|#2)}
\newcommand*{\FXY}{\FF\xv\yv}

\newcommand*{\lap}{\Delta}
\newcommand*{\isp}{\,}

\renewcommand*{\d}[1]{d #1}
\newcommand*{\D}[1]{d #1\isp}
\newcommand{\inta}[1]{\int\D{#1}}
\newcommand{\intb}[2]{\int\d{#1}\D{#2}}
\newcommand{\intin}{\int_0^\infty}
\newcommand{\intoi}{\int_0^1}

\newcommand*{\sums}{\sum_\spin}

\newcommand{\gavg}[2]{\bigl<#2\bigr>_{#1}}
\newcommand{\gAvg}[2]{\Bigl<#2\Bigr>_{#1}}

\newcommand{\scalth}[3]{\langle#1|#2|#3\rangle}
\newcommand{\scal}[2]{\big(#1,#2\big)}

\newcommand{\xa}{\xv}
\newcommand{\xb}{\yv}
\newcommand{\ta}{\tau_x}
\newcommand{\tb}{\tau_y}
\newcommand{\FAB}{\FF\xa\xb}
\newcommand{\ka}{\kv_1}\newcommand{\kb}{\kv_2}
\newcommand{\pa}{\pv_1}\newcommand{\pb}{\pv_2}
\newcommand{\qa}{\qv_1}\newcommand{\qb}{\qv_2}
\newcommand{\ra}{\rv_1}\newcommand{\rb}{\rv_2}
\newcommand{\etal}{\emph{et.\ al.}}
\newcommand{\sC}[2]{\C_{#1#2}}
\newcommand{\sK}[2]{\K_{#1#2}}
\newcommand{\sECRhf}{\srv{E}_\textsc{c}^\textsc{hf}}
\DeclareMathOperator{\Trace}{Tr}
\DeclareMathOperator{\trace}{tr}
\DeclareMathOperator{\Det}{Det}

\begin{document}
%
%
%
%
\newboolean{TC}
\setboolean{TC}{false}
\newcommand{\ITE}[2]{\ifthenelse{\boolean{TC}}{#1}{#2}}
\newcommand{\TCUT}[2]{\ITE{%
	\begin{multline}#1\\\times#2\end{multline}
	}{%
	\begin{equation}#1#2\end{equation}}%
	}
\newcommand{\OCUT}[2]{\ITE{%
	\begin{multline}#1\\#2\end{multline}
	}{%
	\begin{equation}#1#2\end{equation}}%
	}
%
%
%
%
\title{Atoms and Quantum Dots With a Large Number of Electrons: the Ground State Energy}
\author{Herv\'e Kunz}
\email{herve.kunz@epfl.ch}
\author{Rico Rueedi}
\email{rico.rueedi@a3.epfl.ch}
\affiliation{Institute of Theoretical Physics, Ecole Polytechnique F\'ed\'erale de Lausanne (EPFL), CH-1015 Lausanne, Switzerland}
%
%
%
%
\begin{abstract}
We compute the ground state energy of atoms and quantum dots with a large number $N$ of electrons. Both systems are described by a non-relativistic Hamiltonian of electrons in a $d$-dimensional space. The electrons interact via the Coulomb potential. In the case of atoms ($d=3$), the electrons are attracted by the nucleus, via the Coulomb potential. In the case of quantum dots ($d=2$), the electrons are confined by an external potential, whose shape can be varied. We show that the dominant terms of the ground state energy are those given by a semiclassical Hartree-exchange energy, whose $N\to\infty$ limit corresponds to Thomas-Fermi theory. This semiclassical Hartree-exchange theory creates oscillations in the ground state energy as a function of $N$. These oscillations reflect the dynamics of a classical particle moving in the presence of the Thomas-Fermi potential. The dynamics is regular for atoms and some dots, but in general in the case of dots, the motion contains a chaotic component. We compute the correlation effects. They appear at the order $N\ln N$ for atoms, in agreement with available data. For dots, they appear at the order $N$.
\end{abstract}
%
%
%
%
\maketitle
%
%
%
%
\section{Introduction}
Ever since the invention by Thomas \cite{Thomas-PCPS23} and Fermi \cite{Fermi-RANL6} of a simplified theory of an atom with a large number $N$ of electrons, many efforts have been made to systematically improve on it \cite{Morgan-1996}. Corrections were made, then refined, by Dirac \cite{Dirac-PCPS26}, Scott \cite{Scott-PM43}, Schwinger \cite{Schwinger-PRA22,*Schwinger-PRA24}, and Englert and Schwinger \cite{EnglertSchwinger-PRA26, *EnglertSchwinger-PRA29a, *EnglertSchwinger-PRA29b, *EnglertSchwinger-PRA29c}, which add terms of order $\smash{N^{6/3}}$ and $\smash{N^{5/3}}$ to the $\smash{N^{7/3}}$ Thomas-Fermi (TF) energy. TF theory and its corrections, collectively referred to as the statistical atom, thus seemed to result in an expansion of the atomic ground state energy in $\smash{N^{-1/3}}$. After Lieb and Simon \cite{LiebSimon-PRL31,*LiebSimon-AdvMath23} proved that TF theory becomes exact when $N\to\infty$, this expansion was put on a rigorous basis by Fefferman and Seco \cite{FeffermanSeco-AdvMath107a,*FeffermanSeco-HPA71}, who proved that the energy of the statistical atom and the energy of Hartree-Fock (HF) theory are equivalent and exact up to order $\smash{N^{5/3}}$. A crucial step further was made by Schwinger and Englert \cite{EnglertSchwinger-PRA32c}, who showed that there exist oscillating corrections to the ground state energy of order $\smash{N^{4/3}}$ and period $\smash{N^{1/3}}$. They interpreted such corrections as indicating shell effects.

It remains to determine, however, at which order in $N$ both the statistical atom and HF theory break down, and to compute the dominant correlation effects, which are ignored in both approaches. It is this task that we undertake in this paper. 

But we are also interested by the same type of problem in the case of quantum dots. We consider in this paper quantum dots to be $2$-dimensional artificial atoms, whose $N$ electrons are subject to a confining potential and interact by way of the standard $3$-dimensional Coulomb interaction. The determination of the ground state energy as a function of $N$ is of great interest especially since, for a class of confining potentials, the energy has become experimentally accessible \cite{KouwenhovenAustingTarucha-RPP64,*ReimannManninen-RMP74}. The analogy between atoms and quantum dots became quite clear when Lieb, Solovej, and Yngvason \cite{LiebSolovejYngvason-PRB51} proved that in the case of dots also, TF theory becomes exact when $N\to\infty$. We are therefore interested in this case also by the corrections to TF theory. An important difference with the atomic case is that the confining potential isn't necessarily rotationally symmetric. This leads to shell effects sensitive to the nature of the dynamics of a classical particle moving in the presence of the TF potential. Depending on the chosen confining potential, this dynamics can be fully regular, fully chaotic, or, most frequently, mixed. Therefore, quantum dots are ideal laboratories for the study of quantum chaos.

To determine the ground state energy of large atoms and dots, our main idea comes from the fact that after a simple $Z$-dependent rescaling of the coordinates ---where $Z$ is the number of protons in the case of atoms, and $N=Z$ in the case of dots--- the Coulomb interaction between the electrons becomes a weak and long range interaction, indicating the validity of a mean field theory when $N,Z\to\infty$ \cite{LebowitzPenrose-JMP7,*Lieb-JMP7}. But at the same time, the role of the Planck constant is played by a parameter $\smash{\smp=Z^{-1/d}}$, $d$ being the dimension of space. Therefore, large values of $Z$ correspond also to a semiclassical limit. In our case, the mean field theory is simply Hartree theory, and its semiclassical limit is TF theory, in agreement with the theorem of Lieb and Simon \cite{LiebSimon-AdvMath23}.

In order to go beyond TF theory, our strategy is the following. Considering first $\smp$ as an independent parameter, we derive an asymptotic expansion in $\smash{Z\inv}$ of the energy. The dominant term of this expansion is Hartree theory, and the corrections to it correspond to exchange and correlation effects. But remembering that in our case, $\smp$ is small, we then take the semiclassical limit of each term in this expansion. In this way, we have computed the correlation contributions to dominant order. In the case of atoms, they give a term $aZ\ln Z+bZ$, and in the case of dots a term $cZ$. The other contributions to the energy correspond to a Hartree-exchange (HX) theory, which coincides with HF theory up to a certain order in $Z$ (at least $\smash{Z^{5/3}}$ in the case of atoms).

In the case of neutral atoms ($Z=N$), we compare our results for the correlation energy with experimental and numerical values. It was suggested \cite{Englert-1988,ClementiCorongiu-IJQC62} on the basis of these values that the dominant term of the correlation energy is of the order $\smash{N^{4/3}}$, contrary to our results. But we can see that our logarithmic correction allows to fit well the data.

In the case of dots, we completely determine the smooth part of the HX energy to the order $N$. The oscillating part is less important than in the case of atoms, and its analysis is deferred to another article. In the case of atoms, after summarizing known results, which come from the HX energy, we indicate what remains to computed for this part of the energy. This is a delicate problem in semiclassical physics that we have not solved. Our results show that the Schwinger-Englert oscillations can be interpreted as resulting from a trace formula for an integrable system. The integrability in this case is due to the rotational symmetry of the TF potential.
%
%
%
%
\section{The Model For Atoms and Quantum Dots}
We use the dimensionless Hamiltonian
\begin{equation}
\srv H=
-\sum_{\ia=1}^N\srv\lap_\ia
+Z\sum_{\ia=1}^N\V(\srv\xv_\ia)
+\half\sum_{\ia\neq\ib}^N\frac{1}{\norm{\srv\xv_\ia-\srv\xv_\ib}}
\end{equation}
to describe an atom or quantum dot containing $N$ electrons. In the case of atoms, we've used half the Bohr radius for the unit of length and two hartrees for the unit of energy; a similar choice can be made for dots. In the case of atoms, $Z$ denotes the number of protons and $\V(\srv\xv)=-\norm{\srv\xv}\inv$. We have neglected relativistic effects. In the case of dots, $Z$ will be identified with $N$, and $\V(\srv\xv)$ is a confining potential whose form is not specified. At this stage, $\V(\srv\xv)$ is independent of $N$, but if we describe a specific experiment, we may have to consider a smooth dependence of $\V(\srv\xv)$ on $N$.

After the rescaling of the coordinates $\srv\xv=Z^{2/d-1}\xv$, where $d$ is the space dimension, with $d=3$ for atoms and $d=2$ for dots, the Hamiltonian $\srv H$ becomes $\srv H=Z^{2-2/d}H$, where the new Hamiltonian $H$ is given by
\begin{equation}
H =
-\smp^2\sum_{\ia=1}^N\lap_\ia
+\sum_{\ia=1}^N\V(\xv_\ia)
+\frac{1}{2Z}\sum_{\ia\neq\ib}^N
\frac{1}{\norm{\xv_\ia-\xv_\ib}}.
\label{def:H}
\end{equation}
Accordingly, the ground state energies $\sEQ$ of $\srv H$ and $\EQ$ of $H$ are related by $\sEQ= Z^{2-2/d}\EQ$. In $H$, $\smp\doteq Z^{-1/d}$ plays the role of $\hbar$, and the Coulomb interaction between the electrons looks like a mean field type interaction when $N\to\infty$, considering that $\lim_{N\to\infty}Z/N=1$.  Taking first $\smp$ to be an independent parameter, we will give an exact formula for the ground state energy $\EQ$ as an expansion in the small parameter $Z\inv$. This formula decomposes $E$ into two parts, the first corresponding to the HX energy, the second to the correlation energy. We then evaluate the two parts for small $\smp$, determining first the leading order of the correlation energy, then considering the expansion in $\smp$ of the HX energy up to this leading order.
%
%
%
%
\section{Grand-Canonical Formulation}
To determine the ground state energy $E$, we start with the grand-canonical partition function $\Q(\cp,\itemp)$, with $\cp$ the chemical potential, and $\itemp$ the inverse temperature. From $\Q(\cp,\itemp)$, we can get the ground state pressure $\PP(\cp)$ as
\begin{equation}
\PP(\cp)=
\lim_{\itemp\to\infty}
\frac{1}{\itemp}\ln\Q(\cp,\itemp).
\end{equation}
If we define $\EQ\fcp$ by
\begin{equation}
\EQ\fcp\doteq
\cp N-\PP(\cp),
\end{equation}
then the ground state energy at a given number $N$ of electrons is given by
\begin{equation}
\EQ=\inf_\cp\EQ^\cp.
\end{equation}
In the case of atoms, the presence of a continuous spectrum of $H$ would give an infinite $\Q(\itemp,\cp)$, but the problem is easily solved by adding to $\V\FX$ a confining potential, suppressed after the limit $\itemp\to0$ is taken. For clarity of notation, we don't explicitly write this additional confining potential.

\ITE{
\begin{widetext}
Using coherent state representation, we can write the grand-canonical partition function as \cite{NegeleOrland-1988}
\begin{multline}
\Q=
\int\feyn{\psic,\psia}
\exp-\sums
\int\d\xv
\int_0^\itemp\D{\tau}
\psics{\spin}(\xv,\tau)
\big[\partial_\tau-\cp+h'\big]
\psias{\spin}(\xv,\tau)
\\\times
\exp-\frac{1}{2Z}
\int\frac{\d\xv\d\yv}{\norm{\xv-\yv}}
\int_0^\itemp\D\tau
\big[\sums
\psics{\spin}(\xv,\tau)
\psias{\spin}(\xv,\tau)
\big]\big[\sums
\psics{\spin}(\yv,\tau)
\psias{\spin}(\yv,\tau)
\big],
\end{multline}
\end{widetext}
}{
Using coherent state representation, we can write the grand-canonical partition function as \cite{NegeleOrland-1988}
\begin{multline}
\Q=
\int\feyn{\psic,\psia}
\exp-\sums
\int\d\xv
\int_0^\itemp\D{\tau}
\psics{\spin}(\xv,\tau)
\big[\partial_\tau-\cp+h'\big]
\psias{\spin}(\xv,\tau)
\\\times
\exp-\frac{1}{2Z}
\int\frac{\d\xv\d\yv}{\norm{\xv-\yv}}
\int_0^\itemp\D\tau
\big[\sums
\psics{\spin}(\xv,\tau)
\psias{\spin}(\xv,\tau)
\big]\big[\sums
\psics{\spin}(\yv,\tau)
\psias{\spin}(\yv,\tau)
\big],
\end{multline}
}
$\psias\spin(\xv,\tau)$, $\psics\spin(\xv,\tau)$ being Grassmann variables and $h'$ the one-body Hamiltonian $h'\doteq-\smp^2\lap+\V$. By applying a Hubbard-Stratonovich transformation on the Coulomb interaction in this integral, we can integrate over the Grassmann variables, so that the partition function becomes
\begin{equation}
\Q=
\gAvg{\phi;\C\inv}{
\Det^2\big[\partial_\tau-\cp+h'+iZ^{-1/2}\phi\big]},
\end{equation}
$\phi(\xv,\tau)$ being a Gaussian field of zero mean and covariance $\C\inv\FF{\xa,\ta}{\xb,\tb}$, $\C$ being the operator of kernel $\C\FF{\xa,\ta}{\xb,\tb}\doteq\ks{\ta-\tb}\norm{\xa-\xb}\inv$. Here, and in what follows, we denote the determinant and trace $\Det$ and $\Trace$ if they operate on both space and time, and $\det$ and $\trace$ if they operate on space only. 

Let us now make the shift
\begin{equation}
\phi(\xv,\tau)=-iZ^{1/2}\w\FX+\theta(\xv,\tau).
\end{equation}
The partition function can then be written as
\ITE{%
\begin{multline}
\Q=\Q_0
\gAvg{\theta;\C\inv}
{\exp\big[iZ^{1/2}\scal{\C\inv\w}{\theta}\big]
\\
\Det^2\big[1+iZ^{1/2}\K\theta\big]},
\end{multline}
}{%
\begin{equation}
\Q=\Q_0
\gAvg{\theta;\C\inv}
{\exp\big[iZ^{1/2}\scal{\C\inv\w}{\theta}\big]
\Det^2\big[1+iZ^{1/2}\K\theta\big]},
\end{equation}
}
where
\begin{equation}
\Q_0\doteq
\Det^2\big[\partial_\tau-\cp+h\big]
\exp\Big[\half Z\scal{\w}{\C\inv\w}\Big],
\end{equation}
$h$ being the new one-body Hamiltonian $h\doteq h'+\w$, and $\K$ the operator $\smash{\K\doteq[\partial_\tau-\cp+h]\inv}$. On the other hand,
\ITE{%
\begin{multline}
\Det^2\big[
1+iZ^{-1/2}\K\theta\big]=
\\
\exp-2\sum_{n=1}^\infty
\frac{(-i)^n}{nZ^{n/2}}\Trace(\K\theta)^n.
\end{multline}
}{%
\begin{equation}
\Det^2\big[
1+iZ^{-1/2}\K\theta\big]=
\exp-2\sum_{n=1}^\infty
\frac{(-i)^n}{nZ^{n/2}}\Trace(\K\theta)^n.
\end{equation}
}
We can find a $\w\FX$ such that
\begin{equation}
iZ^{1/2}\scal{\C\inv\w}{\theta}+2iZ^{-1/2}\Trace\K\theta=0,
\end{equation}
so that the linear term in $\theta$ disappears. Indeed,
\begin{equation}
\w\FX\doteq
\int\frac{\d\yv}{\norm{\xv-\yv}}\deno\FY,
\label{DF:w}
\end{equation}
where
\begin{equation}
\deno\FX\doteq
\frac{2}{Z}\nf\FF\xv\xv,
\label{DF:deno}
\end{equation}
with
\begin{equation}
\nf\FXY\doteq
\scal{\xv}{\frac{e^{\itemp(\cp-h)}}{1+e^{\itemp(\cp-h)}}\yv},
\end{equation}
does the job, since $-\K\FF{\xv,\tau}{\xv,\tau}=\nf\FF\xv\xv$.

If we introduce the operator $\Gamma$ of kernel
\begin{equation}
\Gamma\FF{\xa,\ta}{\xb,\tb}\doteq2\K\FF{\xa,\ta}{\xb,\tb}\K\FF{\xb,\tb}{\xa,\ta},
\end{equation}
we see that $2\Trace(\K\theta)^2=\scal{\theta}{\Gamma\theta}$, and we can write the partition function in the form
\begin{equation}
\Q=
\Q_0\Q_1
\gAvg{\theta;\C\inv-\Gamma/Z}
{\exp A(\theta)},
\label{ga1}
\end{equation}
where
\begin{equation}
\Q_1\doteq
\gAvg{\theta;\C\inv}
{\exp\frac{1}{2Z}\scal{\theta}{\Gamma\theta}},
\label{ga2}
\end{equation}
and
\begin{equation}
A(\theta)\doteq
-2\sum_{n=3}^\infty
\frac{(-i)^n}{nZ^{n/2}}\trace(\K\theta)^n,
\end{equation}
$\theta$ being a new Gaussian field of zero mean and covariance $\C\inv-\Gamma/Z$ in \eqref{ga1} and $\C\inv$ in \eqref{ga2}. But $\Q_1$ is simply
\ITE{%
\begin{align}
\Q_1&=
\big[\Det\big(1-\C\Gamma/Z)\big]^{-1/2}
\\
&=\exp\half\sum_{n=1}^\infty\frac{1}{nZ^n}\Trace(\C\Gamma)^n.
\end{align}
}{%
\begin{equation}
\Q_1=
\big[\Det\big(1-\C\Gamma/Z)\big]^{-1/2}
=\exp\half\sum_{n=1}^\infty\frac{1}{nZ^n}\Trace(\C\Gamma)^n.
\end{equation}
}
We therefore decompose the pressure $\PP(\cp)$ into four terms as
\begin{equation}
\PP=\PSC+\PXC+\PCA+\PCB,
\label{P:decomposition}
\end{equation}
where
\begin{equation}
\PSC=
\GSL\frac{1}{\itemp}\ln\Q_0,
\end{equation}
is the pressure in a mean field approximation, $\w\FX$ being the mean field. The remaining terms
\begin{equation}
\PXC=
\GSL\frac{1}{2\itemp Z}\Trace(\C\Gamma),
\end{equation}
\begin{equation}
\PCA=
\GSL\frac{1}{2\itemp}\sum_{n=2}^\infty
\frac{1}{nZ^n}\Trace(\C\Gamma)^n,
\end{equation}
and
\begin{equation}
\PCB=
\GSL\frac{1}{\itemp}
\ln\gAvg{\theta;\C\inv-\Gamma/Z}{\exp A(\theta)}.
\end{equation}
correspond to fluctuation effects around the mean field. This decomposition of the pressure will correspond to a natural one for the ground state energy, and the indices \textsc{h}, \textsc{x}, and \textsc{c} foreshadow the nature of the corresponding contributions to the energy.
%
%
%
%
\section{Ground State Energy: Hartree-Exchange and Correlation Decomposition}
In correspondence with the decomposition \eqref{P:decomposition} of the pressure, we decompose the ground state energy as
\begin{equation}
\EQ\fcp=\EH\fcp+\EXC\fcp+\ECA\fcp+\ECB\fcp.
\end{equation}
$\EH\fcp$ is given by
\ITE{
\begin{multline}
\EH\fcp\doteq
\cp N-\PSC
=
\cp N-2\trace\big[(\cp-h)\hs{\cp-h}\big]
\\-
\frac{Z}{2}
\int\frac{\d\xv\d\yv}{\norm{\xv-\yv}}\deno\FX\deno\FY,
\label{EH}
\end{multline}
}{
\begin{equation}
\EH\fcp\doteq
\cp N-\PSC
=
\cp N-2\trace\big[(\cp-h)\hs{\cp-h}\big]
-\frac{Z}{2}
\int\frac{\d\xv\d\yv}{\norm{\xv-\yv}}\deno\FX\deno\FY,
\label{EH}
\end{equation}
}
where we have used equation \eqref{DF:w} which defines $\w\FX$, and where $\deno\FX$ is given by \eqref{DF:deno}, but with both $\w\FX$ and $\deno\FX$ now obtained from the density matrix $\nf\FXY$ in the ground state $\nf\FXY=\scalth\xv{\smhs{\cp-h}}\yv$. We can write $\EH\fcp$ in the form
\ITE{%
\begin{multline}
\EH\fcp=
\cp N-2\int^\cp\D{e} \idos(e)
\\
-\frac{Z}{2}\int\frac{\d\xv\d\yv}{\norm{\xv-\yv}}\deno\FX\deno\FY,
\label{EH2}
\end{multline}
}{
\begin{equation}
\EH\fcp=
\cp N-2\int^\cp\D{e} \idos(e)
-\frac{Z}{2}\int\frac{\d\xv\d\yv}{\norm{\xv-\yv}}\deno\FX\deno\FY,
\label{EH2}
\end{equation}
}
$\idos(e)$ being the integrated density of states $\idos(e)\doteq\trace\smhs{e-h}$ of the Hamiltonian $h$, which we now write as $h=-\smp^2\lap+\W\FX,$ where
\begin{equation}
\W\FX\doteq
\V\FX+\w\FX=
\V\FX+
\int\frac{\d\yv}{\norm{\xv-\yv}}\deno\FY
\label{DEF:W}
\end{equation}
is the self-consistent potential.

$\EXC\fcp$ is given by
\begin{equation}
\EXC\fcp\doteq-\PXC=
-\frac1Z
\int\frac{\d\xv\d\yv}{\norm{\xv-\yv}}
\nf^2\FXY,
\end{equation}
and describes exchange effects. $\ECA\fcp$ and $\ECB\fcp$ describe correlation effects. For $\ECA\fcp\doteq-\PCA$, it is useful to introduce the representation of the kernel of $\K$
\begin{equation}
\K\FF{\xa,\ta}{\xb,\tb}=
\frac{1}{\itemp}\sum_\omega
\K_\omega\FF{\xa}{\xb}e^{i\omega(\ta-\tb)},
\end{equation}
where 
\begin{equation}
\K_\omega\doteq
(i\omega-\cp+h)\inv,
\label{K:frequency:rep}
\end{equation}
$\omega$ being the Matsubara frequencies $\omega=\pi(2n+1)/\itemp$, for $n\in\bbZ$. Then if $\ta\neq\tb$, we have the representation of the kernel of $\Gamma$
\begin{equation}
\Gamma\FF{\xa,\ta}{\xb,\tb}=
\frac2\itemp\sum_\Omega
\Gamma_\Omega\FF{\xa}{\xb}e^{i\Omega(\ta-\tb)},
\end{equation}
with
\begin{equation}
\Gamma_\Omega\FF\xa\xb=
\frac{1}{\itemp}\sum_\omega
\K_\omega\FF\xa\xb\K_{\omega+\Omega}\FF\xb\xa.
\label{DF:gamma:freq}
\end{equation}
In this way, we see that
\begin{equation}
\frac{1}{\itemp}\Trace(\C\Gamma)^n=
\frac{1}{\itemp}\sum_\Omega\trace(\C\Gamma_\Omega)^n,
\label{DF:trace:freq}
\end{equation}
where in the right hand side of this equation, the kernels of the operators $\C$ and $\Gamma_\Omega$, and the trace are defined on $\bbR^d$. The limit $\itemp\to\infty$ is simply taken by replacing $\smash{\frac{1}{\itemp}\sum_\Omega}$ in \eqref{DF:trace:freq} and $\smash{\frac1\itemp\sum_\omega}$ in \eqref{DF:gamma:freq} by $\smash{\int\frac{\d\Omega}{2\pi}}$ and $\smash{\int\frac{\d\omega}{2\pi}}$, respectively. Consequently,
\begin{equation}
\ECA\fcp=
-\frac{1}{2}\sum_{n=2}^\infty
\frac{1}{nZ^n}\int\frac{\d\Omega}{2\pi}\trace(\C\Gamma_\Omega)^n.
\label{ECA:1}
\end{equation}
Finally,
\begin{equation}
\ECB\fcp\doteq-\PCB=
\GSL-\frac{1}{\itemp}
\ln\gavg{\theta;\C\inv-\Gamma/Z}
{\exp A(\theta)}.
\end{equation}

Up to this point the expressions given for the different terms of the energy compose an exact asymptotic expansion in $Z\inv$ for $\EQ\fcp$. From now on, we'll consider only those terms which will contribute, in the subsequent semiclassical limit $\smp\to0$, to the dominant correlation energy. For this purpose, whereas we'll need to keep all terms in $\ECA\fcp$ because of the subtlety of its semiclassical limit, we'll only need to keep $\ECB\fcp$ up to order $Z^{-3}$ (with $\smp$ fixed). If $a_n\doteq-2(-i)^n/n$ and $A_n(\theta)\doteq\trace(\K\theta)^n$, then
\begin{equation}
A(\theta)=
\sum_{n=3}^\infty
\frac{a_n}{Z^{n/2}}A_n(\theta),
\end{equation}
and we write
\ITE{%
\begin{multline}
-\ECB\fcp=
\GSL\frac{1}{\itemp}
\bigg[
\frac{a_4}{Z^2}
\gavg{\theta;\C\inv}{A_4}
+
\frac{a_3^2}{2Z^3}
\gavg{\theta;\C\inv}{A_3^2}
\\
+\frac{a_6}{Z^3}\gavg{\theta;\C\inv}{A_6}
+\frac{a_4B}{Z^3}\bigg]+\ord{Z^{-4}},
\label{EC2}
\end{multline}
}{
\begin{equation}
-\ECB\fcp=
\GSL\frac{1}{\itemp}
\bigg[
\frac{a_4}{Z^2}
\gavg{\theta;\C\inv}{A_4}
+
\frac{a_3^2}{2Z^3}
\gavg{\theta;\C\inv}{A_3^2}
+\frac{a_6}{Z^3}\gavg{\theta;\C\inv}{A_6}
+\frac{a_4B}{Z^3}\bigg]+\ord{Z^{-4}},
\label{EC2}
\end{equation}
}
where
\begin{equation}
B=
\lim_{Z\to\infty}
Z\big[
\gavg{\theta;\C\inv-\Gamma/Z}{A_4}
-\gavg{\theta;\C\inv}{A_4}\big].
\end{equation}
%
%
%
%
%
\section{Fixing the Chemical Potential}
There exists a yet unknown parameter in the expressions of the energies, namely the chemical potential $\cp$. We therefore need to determine the dependence of $\cp$ on $N$. We can write $\EQ^\cp$ in the form
\begin{equation}
\EQ\fcp=
\cp N
-\PP_0(\cp)
-\frac{1}{Z}\PP_1(\cp)
-\sum_{n=2}^\infty
\frac{1}{Z^n}\PP_n(\cp),
\end{equation}
where
\begin{equation}
\PP_0(\cp)=
2\int^\cp\D{e}\idos(e)
+\frac{Z}{2}
\int\frac{\d\xv\d\yv}{\norm{\xv-\yv}}\deno\FX\deno\FY,
\end{equation}
and
\begin{equation}
\PP_1(\cp)=
\int\frac{\d\xv\d\yv}{\norm{\xv-\yv}}\nf^2\FXY,
\end{equation}
the terms $\PP_n(\cp)$ describing correlation effects when $n\geq2$.

The chemical potential is fixed by the equation
\begin{equation}
N=\sum_{n=0}^\infty
\frac{1}{Z^n}\PP'_n(\cp).
\end{equation}
Writing the chemical potential in a $Z\inv$ expansion as
\begin{equation}
\cp=\sum_{n=0}^\infty\frac{\cp_n}{Z^n},
\end{equation}
and the energy as
\begin{equation}
E=\sum_{n=0}^\infty\frac{E_n}{Z^n},
\end{equation}
and keeping only the terms necessary to determine the energy up to the leading order of the correlation energy, we get
\ITE{
\begin{align}
E_0&=
\cp_0N-\PP_0(\cp_0),
\\
E_1&=
\cp_1N-\cp_1\PP_0'(\cp_0)-\PP_1(\cp_0),
\\
\begin{split}
E_2&=
\cp_2N-\cp_2\PP_0'(\cp_0)-\half\cp_1^2\PP_1''(\cp_0)
\\
&\qquad\qquad\qquad-\cp_1\PP_1'(\cp_0)-\PP_2(\cp_0).
\end{split}
\end{align}
}{
\begin{align}
E_0&=
\cp_0N-\PP_0(\cp_0),
\\
E_1&=
\cp_1N-\cp_1\PP_0'(\cp_0)-\PP_1(\cp_0),
\\
E_2&=
\cp_2N-\cp_2\PP_0'(\cp_0)-\half\cp_1^2\PP_1''(\cp_0)
-\cp_1\PP_1'(\cp_0)-\PP_2(\cp_0).
\end{align}
}
If we fix $\cp_0$ by the constraint
\begin{equation}
N=\PP_0'(\cp_0),
\label{CP0}
\end{equation}
we see that 
\begin{equation}
\cp_1=
-\frac{
\PP_1'(\cp_0)
}{
\PP_0''(\cp_0)},
\end{equation}
and therefore
\begin{align}
E_0&=
\cp_0N-\PP_0(\cp_0),\\
E_1&=
-\PP_1(\cp_0),\\
E_2&=
\half\frac{
\big[\PP_1'(\cp_0)\big]^2
}{\PP_0''(\cp_0)}
-\PP_2(\cp_0).
\end{align}
Let us note that equation \eqref{CP0} can be written in the more natural form
\begin{equation}
N=2\idos(\cp_0).
\label{DF:cp0}
\end{equation}
Indeed,
\begin{equation}
\frac{d}{d\cp}
\int^\cp\D{e}\idos(e)=
\idos(\cp)-\trace\big[\hs{\cp-h}\partial_\cp\W\big],
\end{equation}
and
\ITE{
\begin{multline}
\partial_\cp
\int\frac{\d\xv\d\yv}{\norm{\xv-\yv}}
\deno\FX\deno\FY=
\\
2\int\frac{\d\xv\d\yv}{\norm{\xv-\yv}}
\deno\FX\partial_\cp\deno\FY,
\end{multline}
}{%
\begin{equation}
\partial_\cp
\int\frac{\d\xv\d\yv}{\norm{\xv-\yv}}
\deno\FX\deno\FY=
2\int\frac{\d\xv\d\yv}{\norm{\xv-\yv}}
\deno\FX\partial_\cp\deno\FY,
\end{equation}
}
but from equation \eqref{DEF:W},
\begin{equation}
\partial_\cp\W\FX=
\int\frac{\d\xv\d\yv}{\norm{\xv-\yv}}
\partial_\cp\deno\FY,
\end{equation}
so that $\PP_0'(\cp)=2\idos(\cp)$.

We see now that the energy $\EH$, defined by
\begin{equation}
\EH\doteq
\EH\fcp\Big|_{\cp=\cp_0}
\end{equation}
corresponds to the Hartree energy, because equation \eqref{DF:cp0} fixing $\cp_0$ corresponds to the normalization condition
\begin{equation}
\inta\xv\deno\FX=
\frac NZ,
\label{den:normal}
\end{equation}
of the electronic density $\deno\FX$ associated to $h$. 

We can now write the ground state energy as
\begin{equation}
\EQ=
\EHX+
\ECR,
\end{equation}
where $\EHX$ is the Hartree-exchange (HX) energy 
\begin{equation}
\EHX\doteq\EH+\EXC+\ECP,
\end{equation}
with the exchange energy
\begin{equation}
\EXC\doteq
\EXC\fcp\Big|_{\cp=\cp_0},
\end{equation}
and 
\begin{equation}
\ECP\doteq\frac14\frac{[\partial_\cp\EXC\fcp]^2}
{\partial_\cp\idos(\cp)}\bigg|_{\cp=\cp_0},
\end{equation}
and where $\ECR$ is the correlation energy $\ECR\doteq\ECA+\ECB$, with
\begin{equation}
\ECA\doteq
\ECA\fcp\Big|_{\cp=\cp_0},
\end{equation}
and
\begin{equation}
\ECB\doteq
\ECB\fcp\Big|_{\cp=\cp_0}.
\end{equation}
As we will now only need the dominant term $\cp_0$ of the chemical potential, we will denote in what follows $\cp\equiv\cp_0$.
%
%
%
%
\section{Thomas-Fermi Theory}
The TF energy corresponds to the dominant contribution to the ground state energy. It is obtained by keeping the dominant term in $Z\inv$, which is Hartree theory, then taking the semiclassical limit $\smp\to0$. As shown in \eqref{EH2}, the Hartree energy is completely determined by the knowledge of the integrated density of states $\idos(e)$ and the density $\deno\FX$. The semiclassical limit of $\idos(e)$ is given by
\begin{equation}
\idos(e)=
\frac{\alpha_d}{\smp^d}
\inta\xv\hs{e-\W\FX},
\label{SC:idos}
\end{equation}
and, introducing the local chemical potential
\begin{equation}
\cp\FX\doteq\cp-\W\FX,
\end{equation}
and the notation
\begin{equation}
\cp_+\FX\doteq
\cp\FX\hs{\cp\FX},
\end{equation}
the semiclassical limit of $\deno\FX$ is given by
\begin{equation}
\deno\FX=
2\alpha_d\inta\xv
\cp_+^{d/2}\FX,
\label{SC:deno}
\end{equation}
where $\alpha_d\doteq S_d/(d\cd)$, with $\cd=(2\pi)^d$ and $S_d$ the surface of the $d$-dimensional unit sphere $S_d\doteq2\pi^{d/2}/\Gamma(d/2)$. The normalization condition in the Hartree approximation \eqref{den:normal} becomes
\begin{equation}
2\alpha_d\inta\xv\cp_+^{d/2}\FX=
\frac{N}{Z},
\label{cp:normal}
\end{equation}
which results in the self-consistent equation for $\cp\FX$
\begin{equation}
\cp\FX=
\cp-\V\FX
-2\alpha_d\int\frac{\d\yv}{\norm{\xv-\yv}}\cp_+^{d/2}\FY.
\label{SC:sc}
\end{equation}

Combining these results, we get for the semiclassical limit of $\EH$ the TF result, expressed for the original Hamiltonian $\srv H$,
\ITE{
\begin{multline}
\sETF=
\alpha_dZ^{3-d/2}\bigg[\cp+
\inta\xv\cp_+^{d/2}\FX\V\FX
\\
+\frac{d-2}{d+2}\inta\xv\cp_+^{d/2+1}\FX
\bigg].
\end{multline}
}{
\begin{equation}
\sETF=
\alpha_dZ^{3-d/2}\bigg[\cp+
\inta\xv\cp_+^{d/2}\FX\V\FX
+\frac{d-2}{d+2}\inta\xv\cp_+^{d/2+1}\FX
\bigg].
\end{equation}
}
In the case of atoms, $\cp\FX\equiv\cp(|\xv|)$ because of the rotational symmetry of $\V\FX$, and we will use the fact that $\cp_+\FX=\ord{|\xv|\inv}$ when $|\xv|\to0$ and $\cp_+(\xv)=\ord{|\xv|^{-4}}$ when $|\xv|\to\infty$. And we can recall that $\cp=0$ in the case of neutral atoms ($Z=N$). In the case of dots of confinement such that $\V\FX\to\infty$ when $\norm\xv\to\infty$, it has been proven \cite{LiebSolovejYngvason-PRB51} that $\cp\FX$ has compact support and is bounded on this support.
%
%
%
%
\section{Semiclassical Estimate of the Correlation Energy}
We now consider the correlation energy in the semiclassical limit. In particular, this limit corresponds to taking, in $\ECA$ and $\ECB$, for the chemical potential $\cp$ and the self-consistent potential $\W\FX$, their TF values, as defined in the previous sections.
%
%
%
%
\subsection{Energy \textnormal{$\ECA$}}
From \eqref{K:frequency:rep} it follows that
\begin{equation}
\K_\omega\FF{\xa}{\xb}=
\inta{e}\frac{\den_e\FF{\xa}{\xb}}{i\omega-\cp+e},
\end{equation}
where $\den_e\FF\xa\xb=\scalth{\xa}{\smks{e-h}}{\xb}$. Therefore, since
\begin{equation}
\Gamma_\Omega\FF\xa\xb=
2\int\frac{\d\omega}{2\pi}
\K_\omega\FF\xa\xb
\K_{\omega+\Omega}\FF\xa\xb,
\end{equation}
we have
\ITE{
\begin{multline}
\Gamma_\Omega\FF\xa\xb=
\intb{e_1}{e_2}
\frac{\den_{e_1}\FF\xa\xb\den_{e_2}\FF\xb\xa}
{i\Omega+e_1-e_2}
\\\times
\Big[
\hs{e_2-\cp}
-\hs{e_1-\cp}\Big].
\label{GO:1}
\end{multline}
}{
\begin{equation}
\Gamma_\Omega\FF\xa\xb=
\intb{e_1}{e_2}
\frac{\den_{e_1}\FF\xa\xb\den_{e_2}\FF\xb\xa}
{i\Omega+e_1-e_2}
\Big[
\hs{e_2-\cp}
-\hs{e_1-\cp}\Big].
\label{GO:1}
\end{equation}
}
From \eqref{ECA:1}, we see that $\ECA$ is given by
\begin{equation}
\ECA=
-\half\sum_{n=2}^\infty
\frac{1}{n}Y_n,
\end{equation}
where
\begin{equation}
Y_n\doteq
\frac{1}{Z^n}
\int\frac{\d\Omega}{2\pi}
\trace(\C\Gamma_\Omega)^n.
\end{equation}

Let us first consider the dominant term in $Z\inv$, namely $Y_2$, and define $\ECS\doteq-Y_2/4$. We can write $Y_2$ as
\begin{equation}
Y_2=
\frac{1}{Z^2}
\int\frac{\d\Omega}{2\pi}
\intb{\lambda_1}{\lambda_2}
\frac{\trace(G_{\lambda_1}G_{\lambda_2})}
{(i\Omega+\lambda_1)(i\Omega+\lambda_2)},
\end{equation}
where
\begin{multline}
G_\lambda\FF\xa\xb\doteq
2\intb{e_1}{e_2}\inta{\zv}
\C\FF\xa\zv
\den_{e_1}\FF\zv\xb
\den_{e_2}\FF\zv\xb
\\\times
\ks{\lambda-(e_1-e_2)}
\Big[\hs{e_2-\cp}-\hs{e_1-\cp}\Big].
\end{multline}
Then
\begin{equation}
Y_2=-\frac{2}{Z^2}
\intb{\lambda_1}{\lambda_2}
\frac{\hs{\lambda_1}\hs{-\lambda_2}}
{\lambda_1-\lambda_2}
\trace\big(G_{\lambda_1}G_{\lambda_2}\big),
\end{equation}
or, more conveniently,
\begin{equation}
Y_2=
\frac{2}{Z^2}\int_0^\infty\D{t}\trace(L^2_t),
\end{equation}
with
\ITE{
\begin{multline}
L_t\FF\xa{\xb}\doteq
2\intb{e_1}{e_2}\inta{\zv}
\den_{e_1}\FF\zv\xb
\den_{e_2}\FF\zv\xb
\\\times
\C\FF\xa\zv
e^{-t(e_1-e_2)}
\hs{e_1-\cp}\hs{\cp-e_2}.
\end{multline}
}{
\begin{equation}
L_t\FF\xa{\xb}\doteq
2\intb{e_1}{e_2}\inta{\zv}
\den_{e_1}\FF\zv\xb
\den_{e_2}\FF\zv\xb
\C\FF\xa\zv
e^{-t(e_1-e_2)}
\hs{e_1-\cp}\hs{\cp-e_2}.
\end{equation}
}
Inserting the semiclassical expression for $\den_e\FF\zv\xb$
\ITE{
\begin{multline}
\den_e\FF\xa\xb=
\frac{1}{\smp^d\cd}
\\\times
\inta{\pv}
\ks{e-\W\big(\tfrac{\xa+\xb}{2}\big)-\pv^2}
e^{i\pv(\xa-\xb)/\smp},
\end{multline}
}{
\begin{equation}
\den_e\FF\xa\xb=
\frac{1}{\smp^d\cd}
\inta{\pv}
\ks{e-\W\big(\tfrac{\xa+\xb}{2}\big)-\pv^2}
e^{i\pv(\xa-\xb)/\smp},
\end{equation}
}
we get
\ITE{
\begin{multline}
L_t\FF{\xv+\smp\rv/2}{\xv-\smp\rv/2}=
\frac{2}{\smp^{2d}\cd^2}
\\\times
\intb{\qv}{\kv}
\inta{\zv}
\fou\C(\pv-\kv/\smp)
e^{i\xa\qv-i\zv\qv}
\alpha_t\big(\kv;\cp(\zv)\big),
\end{multline}
}{
\begin{equation}
L_t\FF{\xv+\smp\rv/2}{\xv-\smp\rv/2}=
\frac{2}{\smp^{2d}\cd^2}
\intb{\qv}{\kv}
\inta{\zv}
\fou\C(\pv-\kv/\smp)
e^{i\xa\qv-i\zv\qv}
\alpha_t\big(\kv;\cp(\zv)\big),
\end{equation}
}
where $\fou\C(\pv)$ is the Fourier transform of the Coulomb potential
\begin{equation}
\fou\C(\pv)\doteq
\frac{S_d}{\cd}\norm{\pv}^{d-1},
\end{equation}
and
\ITE{
\begin{multline}
\alpha_t(\kv;\cp)\doteq
\intb{\qv_1}{\qv_2}
e^{-t(q_1^2-q_2^2)}
\\\times
\hs{q_1^2-\cp}\hs{\cp-q_2^2}\ks{\kv+\qv_1+\qv_2}.
\end{multline}
}{
\begin{equation}
\alpha_t(\kv;\cp)\doteq
\intb{\qv_1}{\qv_2}
e^{-t(q_1^2-q_2^2)}
\hs{q_1^2-\cp}\hs{\cp-q_2^2}\ks{\kv+\qv_1+\qv_2}.
\end{equation}
}
In this way we obtain the useful representation
\begin{equation}
-\ECS=
\smp^{d-2}\frac{S_d^2}{2\cd^3}X_\smp,
\end{equation}
where
\ITE{
\begin{multline}
X_\smp\doteq
\intb\pv\kv
k\Lambda(\kv,\pv)
\\\times
\norm{\kv+\smp\pv/2}^{1-d}
\norm{\kv-\smp\pv/2}^{1-d},
\label{XE}
\end{multline}
}{
\begin{equation}
X_\smp\doteq
\intb\pv\kv
k\Lambda(\kv,\pv)
\norm{\kv+\smp\pv/2}^{1-d}
\norm{\kv-\smp\pv/2}^{1-d},
\label{XE}
\end{equation}
}
and
\begin{equation}
\Lambda(\kv,\pv)\doteq
\frac{4}{\cd k}
\intin\D{t}
\Big|\inta\xv \alpha_t(\kv;\cp\FX)e^{i\pv\xv}\Big|^2,
\end{equation}
where the $k\inv$ factor is chosen so that $\Lambda$ is finite when $k\to0$.

Let us try to apply the semiclassical limit simply by setting $\smp=0$ in $X_\smp$. We have
\begin{equation}
\inta\pv \Lambda(\kv,\pv)=
\frac{4}{k}\intin\d{t}\inta\xv \alpha_t^2(\kv,\cp\FX)
\end{equation}
and since $\alpha_t$ satisfies the scaling relation
$
\alpha_t(\kv\cp^{1/2},\cp)=
\cp^{d/2}\alpha_{\cp t}(\kv,1),
$
we find, using equation \eqref{cp:normal}, 
\begin{equation}
X_0=
\frac{2d\cd}{S_d}\frac{N}{Z}I_d,
\end{equation}
where
\begin{equation}
I_d\doteq
\intin\d{t}\inta\kv k^{2-2d}
\alpha_t^2(\kv;1).
\end{equation}
In the case of dots, $I_2$ is finite and, after a somewhat lengthy computation \cite{Rueedi-2009}, we can evaluate it to
\begin{equation}
I_2=
2\pi^3(1-\ln2),
\end{equation}
so that with $N=Z$,
\begin{equation}
-\ECS=\half(1-\ln 2)=0.1534.
\end{equation}
It is interesting to note that this result is universal, that is, independent of the confining potential.

In the case of atoms, $\ECS$ corresponds to the first second order contribution, of $r_s$ perturbation theory, to the correlation energy for the homogeneous electron gas. It is logarithmically divergent. For the homogeneous electron gas, a finite result of order $\ln r_s$ for the correlation energy is obtained by summing the most divergent terms of all higher order contributions \cite{Gell-MannBrueckner-PR106}. The atom, however, isn't fully homogeneous, it is effectively confined. This confinement, related to an inhomogeneous chemical potential, translates into an effective cut-off in the integral in \eqref{XE}, which will result in a finite $X_\smp$ of order $\ln\smp\inv$. A lengthy computation (see appendix) gives the result
\begin{equation}
X_\smp=
X_\text{log}\ln\smp\inv
+X_\text{lin}+\ord{\smp},
\end{equation}
where $X_\text{log}=2(1-\ln2)(2\pi)^5N/Z,$ and
\begin{multline}
X_\text{lin}=
\frac{N}{Z}(2\pi)^5
\Big[\frac{23}{6}-\frac{\pi^2}{4}+\frac83\ln2-2G(1-\ln2)-4\ln^22\Big]
\\
+\half(4\pi)^4
\Big[(1-\ln2)A-\frac{1}{2\pi}B\Big],
\label{X0}
\end{multline}
where $G$ is Catalan's constant,
\begin{equation}
A\doteq
\intin\D{r}r^2\cp_+^{3/2}(r)\ln\cp_+^{1/2}(r),
\label{xA}
\end{equation}
and
\begin{equation}
B\doteq
\intin\d{t}\intin\D{p}\ln(p)g^2(p,t),
\label{xB}
\end{equation}
with
\ITE{
\begin{multline}
g(p,t)\doteq
\intin\D{r}r\cp'_+(r)e^{-t\cp_+^{1/2}(r)}
\\\times
\Big[\cos(pr)-\frac{\sin(pr)}{pr}\Big]
\end{multline}
}{
\begin{equation}
g(p,t)\doteq
\intin\D{r}r\cp'_+(r)e^{-t\cp_+^{1/2}(r)}
\Big[\cos(pr)-\frac{\sin(pr)}{pr}\Big]
\end{equation}
}
Note that the logarithmic term is universal, but the linear correction contains non-universal terms.

It remains, however, to consider the remainder
\begin{equation}
R\doteq-\frac{1}{2}\sum_{n=3}\frac{1}{n}Y_n,
\end{equation}
in the semiclassical limit. It is convenient to rewrite $R$ as
\begin{equation}
R=-\half
\sum_{n=3}^\infty
\frac{(-1)^n}{n}
\int\frac{\d\Omega}{2\pi}\smp 
X_n^\smp(\Omega),
\end{equation}
where
\begin{equation}
X_n^\smp(\Omega)\doteq
\trace\Big[-Z\inv\C^{1/2}\Gamma_{\smp\Omega}\C^{1/2}\Big]^n,
\end{equation}
because it easily follows from \eqref{GO:1} that the operator $-Z\inv\C^{1/2}\Gamma_{\smp\Omega}\C^{1/2}$ is self-adjoint and positive definite. We can prove that in the case of dots, $\lim_{\smp\to0}\smp\int\d\Omega X_n^\smp(\Omega)=0$, for $n\geq3$, so that $\ECA=\ECS$. But in the case of atoms, $\smash{\int\d\Omega X_n^0(\Omega)}$ is finite when $n\geq3$.

We now consider exclusively the case of atoms. We can put the semiclassical representation of the kernel of $-Z\inv\C^{1/2}\Gamma_{\smp\Omega}\C^{1/2}$ in the form
\ITE{
\begin{multline}
\big[-Z\inv\C^{1/2}\Gamma_{\smp\Omega}\C^{1/2}\big]\FAB=
\\
\frac{1}{4\pi^5\chi_3}
\intb{\kv_1}{\kv_2}
T_\Omega^\smp\FF{\kv_1}{\kv_2}
e^{i\kv_1\xa-i\kv_2\xb},
\end{multline}
}{
\begin{equation}
\big[-Z\inv\C^{1/2}\Gamma_{\smp\Omega}\C^{1/2}\big]\FAB=
\frac{1}{4\pi^5\chi_3}
\intb{\kv_1}{\kv_2}
T_\Omega^\smp\FF{\kv_1}{\kv_2}
e^{i\kv_1\xa-i\kv_2\xb},
\end{equation}
}
where
\ITE{
\begin{multline}
T_\Omega^\smp\FF{\kv_1}{\kv_2}\doteq
\frac{1}{k_1k_2}
\\\times
\inta\yv
e^{i\yv(\kv_2-\kv_1)}
b_\Omega^\smp(\kv_1+\kv_2;\cp_+\FY),
\end{multline}
}{
\begin{equation}
T_\Omega^\smp\FF{\kv_1}{\kv_2}\doteq
\frac{1}{k_1k_2}
\inta\yv
e^{i\yv(\kv_2-\kv_1)}
b_\Omega^\smp(\kv_1+\kv_2;\cp_+\FY),
\end{equation}
}
and
\ITE{
\begin{multline}
b_\Omega^\smp(\pv;\cp)\doteq
\frac{1}{\smp}
\inta\qv
\frac{(\pv,\qv)}
{\Omega^2+(\pv,\qv)^2}
\\\times
\hs{(\smp\pv/4+\qv)^2-\cp}\hs{\cp-(\smp\pv/4-\qv)^2}.
\end{multline}
}{
\begin{equation}
b_\Omega^\smp(\pv;\cp)\doteq
\frac{1}{\smp}
\inta\qv
\frac{(\pv,\qv)}
{\Omega^2+(\pv,\qv)^2}
\hs{(\smp\pv/4+\qv)^2-\cp}\hs{\cp-(\smp\pv/4-\qv)^2}.
\end{equation}
}
Therefore
\begin{equation}
X_n^\smp(\Omega)=
\frac{1}{(4\pi^5)^n}\trace(T_\Omega^\smp)^n.
\end{equation}
This representation is useful because $b^0_\Omega(\pv;\cp)$, given by
\begin{equation}
b_\Omega^0(\pv;\cp)=
\inta\qv\ks{q^2-\cp}
\frac{(\pv,\qv)^2}{\Omega^2+(\pv,\qv)^2},
\end{equation}
is finite. More explicitely $\smash{b_\Omega^0(\pv;\cp)\doteq\cp^{1/2}b(p\cp^{1/2}/\Omega),}$ where $b(x)=2\pi[1-x\inv\arctan x]$.

It remains to verify that $X_n^0(\Omega)$ is integrable. For this purpose, we will use the inequality $\norm{X_n^0(\Omega)}\leq[X_2^0(\Omega)]^{n/2}$. But
\ITE{
\begin{multline}
X_2^0(\Omega)=
2\intb{\zv_1}{\zv_2}
\int\frac{\d\pv\d\qv}
{(\pv+\qv)^2(\pv-\qv)^2}
\\\times
e^{i\qv(\zv_2-\zv_1)}
\prod_{j=1}^2
b_\Omega^0(\qv;\cp_+(\zv_j)).
\end{multline}
}{
\begin{equation}
X_2^0(\Omega)=
2\intb{\zv_1}{\zv_2}
\int\frac{\d\pv\d\qv}
{(\pv+\qv)^2(\pv-\qv)^2}
e^{i\qv(\zv_2-\zv_1)}
\prod_{j=1}^2
b_\Omega^0(\qv;\cp_+(\zv_j)).
\end{equation}
}
Consider first the case $\Omega\gg1$. Writing $X_2^0(\Omega)$ as
\ITE{
\begin{multline}
X_2^0(\Omega)=
\frac{2}{\Omega}
\int\frac{\d\pv\d\qv}
{(\pv+\qv)^2(\pv-\qv)^2}
\intb\xv\rv
e^{i\qv\rv}
\\\times
b_1^0\big(\pv;\cp_+(\xv+\tfrac{\rv}{2\Omega})\big)
b_1^0\big(\pv;\cp_+(\xv-\tfrac{\rv}{2\Omega})\big),
\end{multline}
}{
\begin{equation}
X_2^0(\Omega)=
\frac{2}{\Omega}
\int\frac{\d\pv\d\qv}
{(\pv+\qv)^2(\pv-\qv)^2}
\intb\xv\rv
e^{i\qv\rv}
b_1^0\big(\pv;\cp_+(\xv+\tfrac{\rv}{2\Omega})\big)
b_1^0\big(\pv;\cp_+(\xv-\tfrac{\rv}{2\Omega})\big),
\end{equation}
}
we see that
\ITE{
\begin{multline}
\lim_{\Omega\to\infty}
\Omega X_2^0(\Omega)=
2\chi_3^2
\Big[\inta\xv\cp_+^{3/2}\FX\Big]
\\\times
\Big[\intin\frac{\d p}{p^2}b^2(p)\Big],
\end{multline}
}{
\begin{equation}
\lim_{\Omega\to\infty}
\Omega X_2^0(\Omega)=
2\chi_3^2
\Big[\inta\xv\cp_+^{3/2}\FX\Big]
\Big[\intin\frac{\d p}{p^2}b^2(p)\Big],
\end{equation}
}
which is finite. Hence $X_2^0(\Omega)=\ord{\Omega\inv}$. In the limit $\Omega\ll1$, we start from the expression
\begin{equation}
X_2^0(\Omega)=
4S_3^3
\intin\frac{\d p}{p^3}G_\Omega(p).
\end{equation}
Here
\ITE{
\begin{multline}
G_\Omega(p)=
\intin\D{x}g(x)
\\\times
\Big[
\intin\D{r}p\cp_+^{1/2}(r)
b\Big(\frac{p\cp_+^{1/2}(r)}{\Omega}\Big)
r\sin(xpr)
\Big]^2,
\end{multline}
}{
\begin{equation}
G_\Omega(p)=
\intin\D{x}g(x)
\Big[
\intin\D{r}p\cp_+^{1/2}(r)
b\Big(\frac{p\cp_+^{1/2}(r)}{\Omega}\Big)
r\sin(xpr)
\Big]^2,
\end{equation}
}
and
\begin{equation}
g(x)=
\frac{1}{x}
\frac{1}{x^2+1}
\ln\frac{x+1}{\norm{x-1}}.
\end{equation}
One can show that
\begin{equation}
\int_0^\Omega\frac{\d p}{p^3}G_\Omega(p)\leq c_1,
\end{equation}
\begin{equation}
\int_1^\infty\frac{\d p}{p^3}G_\Omega(p)\leq c_2,
\end{equation}
and
\begin{equation}
\int_\Omega^1\frac{\d p}{p^3}G_\Omega(p)\leq c_3\ln\Omega\inv,
\end{equation}
if $\Omega\leq\Omega_0$. Consequently, if $\Omega>\Omega_0$, $\norm{X_n^0(\Omega)}\leq
d_1/\Omega^{n/2},$ and if $\Omega\leq\Omega_0$, $\norm{X_n^0(\Omega)}\leq
d_2(\ln\Omega\inv)^{n/2},$ so that $\smash{\int\frac{\d\Omega}{2\pi}X_n^0(\Omega)}$ is finite if $n\geq3$. 

We can now summarize all these results. In the case of atoms, we have
\OCUT{
-\ECA=
0.03109\ \frac{N}{Z^{4/3}}\ln Z^{1/3}
}{
+Z^{-1/3}\big[0.03700+x_a+x_b+x\big],
}
where $x_a=0.01979\ A$, $A$ being defined by \eqref{xA}, $x_b=0.01027\ B$, $B$ being defined by $\eqref{xB}$, and
\begin{equation}
x\doteq
\half\sum_{n=3}^\infty
\frac{(-1)^n}{n}
\int\frac{\d\Omega}{2\pi}
\trace\big(T_\Omega^0\big)^n
\Big(\frac{1}{4(2\pi)^5}\Big)^n,
\end{equation}
$T_\Omega^0$ being the compact operator of kernel
\TCUT{
T_\Omega^0\FF{\kv_1}{\kv_2}\doteq
\frac{1}{k_1k_2}
\inta\zv
\cp_+^{1/2}(\zv)
e^{i\zv(\kv_2-\kv_1)}
}{
b\Big(\frac{\norm{\kv_1+\kv_2}}{\Omega}\cp_+^{1/2}(\zv)\Big),
}
where $b(x)=2\pi[1-x\inv\arctan x]$.
%
%
%
%
\subsection{Energy \textnormal{$\ECB$}}
Starting from \eqref{EC2} and anticipating the result of the semiclassical limit, we decompose $\ECB$ as
\begin{equation}
-\ECB=
2a_4A_d
+a_4B_d
+\frac{9}{2}a_3^2C_d+R_d.
\end{equation}
In order to give a simplified expression for these terms, we introduce the notations $(\xv_j,\tau_j)\equiv j$, $\C(\xv_j,\tau_j|\xv_k,\tau_k)\equiv\sC jk$, and $\K(\xv_j,\tau_j|\xv_k,\tau_k)\equiv\sK jk$, and define $L_{12\ldots n}\doteq\prod_{j=1}^n\sK j{(j+1)}$ with $\sK n{(n+1)}\doteq\sK n1$. Then
\begin{equation}
A_d\doteq
\GSL\frac{1}{\itemp Z^2}
\int\prod_{j=1}^4\D{j}
L_{1234}\sC12\sC34,
\end{equation}
\begin{equation}
B_d\doteq
\GSL\frac{1}{\itemp Z^2}
\int\prod_{j=1}^4\D{j}
L_{1234}\sC13\sC24,
\end{equation}
and
\begin{equation}
C_d\doteq
\GSL\frac{1}{\itemp Z^3}
\intb12
f_1f_2,
\end{equation}
where
\begin{equation}
f_1\doteq
\intb23L_{123}\sC23.
\end{equation}
If
\begin{equation}
R_d\doteq
\frac{3}{2}a_3^2D_d
+a_4E_d
+a_6F_d,
\end{equation}
then
\TCUT{
D_d\doteq
\GSL
\frac{1}{\itemp Z^3}
\int\prod_{j=1}^6\D j
L_{123}L_{456}
\sC14
}{
\Big[\sC25\sC36+
\sC26\sC35\Big],
}
and
\OCUT{
E_d\doteq
\GSL\frac{1}{\itemp Z^3}
\intb12(\C\Gamma\C)_{12}
\bigg[
4\K_{12}(\K\hat\K\K)_{21}
}{
+2\intb34\sK13\sC34\sK41\sK32\sK24\bigg],
}
where $\hat\K_{jk}\doteq\sK jk\sC jk$, and
\begin{equation}
F_d\doteq
\GSL
\frac{1}{\itemp Z^3}
\int\prod_{j=1}^6\D j
L_{123456}
\gAvg{\theta;\C\inv}
{\prod_{k=1}^6\theta_k}.
\end{equation}
The semiclassical computation of $R_d$, which corresponds to terms of order $Z^{-3}$ requires a lot of work, that we do not reproduce here. The final result is that $R_d=\ord{\smp^{2d-3}\ln\smp\inv}$, so that we can ignore $R_d$, since the other terms will be of order $\smp^{d-2}$.

We begin by first giving the expressions for $A_d$, $B_d$, and $C_d$ when the limit $\itemp\to\infty$ is taken. They are
\newcommand{\ea}{e_1}
\newcommand{\eb}{e_2}
\newcommand{\ec}{e_3}
\newcommand{\ed}{e_4}
\renewcommand{\xa}{\xv_1}
\renewcommand{\xb}{\xv_2}
\newcommand{\xc}{\xv_3}
\newcommand{\xd}{\xv_4}
\TCUT{
A_d=
-\frac{2}{Z^2}
\int\frac{\d\ea\d\eb}{\ea-\eb}
\trace\big(
\hat\K\den_{\ea}\hat\K\den_{\eb}\big)
}{
\hs{\ea-\cp}\hs{\cp-\eb},
}
where now $\hat\K\FXY=-\nf\FXY\C\FXY$,
\TCUT{
B_d=
\frac{2}{Z^2}
\int\prod_{j=1}^4\d{e_j}
\frac{L(\ea,\ec;\eb,\ed)}
{\ea+\ec-\eb-\ed}
}{
\hs{\ea-\cp}\hs{\ec-\cp}
\hs{\cp-\eb}\hs{\cp-\ed},
\label{BD}
}
where
\begin{multline}
L(\ea,\ec;\eb,\ed)\doteq
\int\prod_{j=1}^4\D{\xv_j}
\C\FF\xa\xc
\C\FF\xb\xd
\\\times
\den_{\ea}\FF\xa\xb
\den_{\ec}\FF\xc\xd
\den_{\eb}\FF\xb\xc
\den_{\ed}\FF\xd\xa,
\label{L1234}
\end{multline}
and
\begin{equation}
C_d=
\frac{1}{Z^3}
\int\frac{\d\xa\d\xb}{\norm{\xa-\xb}}
f(\xa)f(\xb),
\end{equation}
where
\TCUT{
f(\xv)\doteq
-2\int
\frac{\d\ea\d\eb}
{\ea-\eb}
\big(\den_{\ea}\hat\K\den_{\eb}\big)\FF\xv\xv
}{
\hs{\ea-\cp}\hs{\cp-\eb}.
\label{DF:f}
}
Let us then take the semiclassical limit of these expressions. We have
\TCUT{
-\hat\K\FF\xa\xb=
\frac{1}{\smp^{d+1}\cd}
\inta{\qv}e^{i\qv(\xa-\xb)/\smp}
}{
g_d\big(\qv,\cp\big(\tfrac{\xa+\xb}{2}\big)\big),
}
where
\begin{equation}
g_d(\qv,\cp)\doteq
c_d\int\frac{\d\pv}{p^{d-1}}
\hs{\cp-(\qv-\pv)^2},
\end{equation}
with $c_d\doteq S_d/\cd$, so that
\ITE{
\begin{widetext}
\begin{multline}
\frac{A_d}{\smp^{d-2}}=
-\frac{2}{\cd^4}
\int\d{\qa}\d{\qb}\d{\pa}\d{\pb}\d\xv\d\sv\d\rv\D\tv
g_d\big(\qa+\tfrac{\qb}{2};\cp(\xv+\smp\tfrac{\sv+\rv}{2})\big)
g_d\big(\qa-\tfrac{\qb}{2};\cp(\xv+\smp\tfrac{\sv-\rv}{2})\big)
\\\times
e^{i[-(\qb,\sv)+(\pb,\rv)+(\tv,\qa-\pb)]}
\frac{
\hs{(\pa+\tfrac\pb2)^2-\cp(\xv+\smp\sv)}
\hs{\cp\FX-(\pa-\tfrac\pb2)^2}
}{
2(\qv,\pb)+\cp\FX-\cp(\xv+\smp\sv)
}.
\end{multline}
\end{widetext}
}{
\begin{multline}
\frac{A_d}{\smp^{d-2}}=
-\frac{2}{\cd^4}
\int\d{\qa}\d{\qb}\d{\pa}\d{\pb}\d\xv\d\sv\d\rv\D\tv
g_d\big(\qa+\tfrac{\qb}{2};\cp(\xv+\smp\tfrac{\sv+\rv}{2})\big)
g_d\big(\qa-\tfrac{\qb}{2};\cp(\xv+\smp\tfrac{\sv-\rv}{2})\big)
\\\times
e^{i[-(\qb,\sv)+(\pb,\rv)+(\tv,\qa-\pb)]}
\frac{
\hs{(\pa+\tfrac\pb2)^2-\cp(\xv+\smp\sv)}
\hs{\cp\FX-(\pa-\tfrac\pb2)^2}
}{
2(\qv,\pb)+\cp\FX-\cp(\xv+\smp\sv)
}.
\end{multline}
}
It is now easy to take the limit $\smp\to0$, so that
\TCUT{
\lim_{\smp\to0}\frac{A_d}{\smp^{d-2}}=
-\frac{1}{\cd}
\intb\xv\pv
g_d^2(\pv;\cp\FX)
}{
\ks{\pv^2-\cp(\xv)}.
}
The scaling relation $\cp^{-1/2}g_d(\cp^{1/2}\pv;\cp)=g_d(\pv;1)$ and the normalisation condition \eqref{cp:normal} give
\begin{equation}
A_d=
-\smp^{d-2}\frac{N}{Z}\frac{4d}{S_d^2},
\end{equation}
since $g_d(1;1)=4/S_d$.

\ITE{%
\begin{widetext}
To compute $B_d$, we write $L(\ea,\ec;\eb,\ed)$ semiclassically as
\begin{multline}
L(\ea,\ec;\eb,\ed)=
\frac{c_d^2}{\smp^{d+2}\cd^4}
\int\d\pa\d\pb\d\qa\d\qb\d\kv\d\xv\d\ra\D\rb
\big|\pa+\tfrac{\pb}{2}\big|^{1-d}
\big|\pa-\tfrac{\pb}{2}\big|^{1-d}
e^{i(\ra,\pa+\qb)+i(\rb,\pb-2\qa)}
\\\shoveleft{\phantom{L(\ea,\ec;\eb,\ed)=}\times
\ks{\ea-\big(\qa+\tfrac\kv2\big)^2-\W\big(\xv+\smp\tfrac\ra4\big)}
\ks{\ec-\big(\qa-\tfrac\kv2\big)^2-\W\big(\xv-\smp\tfrac\ra4\big)}
}\\\shoveleft{\phantom{L(\ea,\ec;\eb,\ed)=}\times
\ks{\eb-\big(\qb+\tfrac\kv2\big)^2-\W\big(\xv+\smp\tfrac\rb2\big)}
\ks{\ed-\big(\qb-\tfrac\kv2\big)^2-\W\big(\xv+\smp\tfrac\rb2\big)}.\hfill}
\end{multline}
\end{widetext}
}{%
To compute $B_d$, we write $L(\ea,\ec;\eb,\ed)$ semiclassically as
\begin{multline}
L(\ea,\ec;\eb,\ed)=
\frac{c_d^2}{\smp^{d+2}\cd^4}
\int\d\pa\d\pb\d\qa\d\qb
\int\d\xv\d\ra\D\rb
\big|\pa+\tfrac{\pb}{2}\big|^{1-d}
\big|\pa-\tfrac{\pb}{2}\big|^{1-d}
\\\shoveleft{\phantom{L(\ea,\ec;\eb,\ed)=}\times
e^{i(\ra,\pa+\qb)+i(\rb,\pb-2\qa)}
}\\\shoveleft{\phantom{L(\ea,\ec;\eb,\ed)=}\times
\ks{\ea-\big(\qa+\tfrac\kv2\big)^2-\W\big(\xv+\smp\tfrac\ra4\big)}
\ks{\ec-\big(\qa-\tfrac\kv2\big)^2-\W\big(\xv-\smp\tfrac\ra4\big)}
}\\\times
\ks{\eb-\big(\qb+\tfrac\kv2\big)^2-\W\big(\xv+\smp\tfrac\rb2\big)}
\ks{\ed-\big(\qb-\tfrac\kv2\big)^2-\W\big(\xv+\smp\tfrac\rb2\big)}.
\end{multline}
}
In this form, the limit $\smp\to0$ is easily taken, and we get
\TCUT{
B_d=
\smp^{d-2}\frac{2^dc_d^2}{\cd}
\int\frac{\d\pa\d\pb}{\pa^2-\pb^2}
|\pa+\pb|^{1-d}|\pa-\pb|^{1-d}
}{
\inta\xv
f\big(\pa,\pb;\cp_+\FX\big),
}
where
\begin{multline}
f(\pa,\pb;\cp)\doteq
\inta\qv
\hs{(\qv-\pa)^2-\cp}
\hs{(\qv+\pa)^2-\cp}
\\\times
\hs{\cp-(\qv+\pb)^2}
\hs{\cp-(\qv-\pb)^2}.
\end{multline}
But $\cp^{-d/2}f(\cp^{1/2}\pa,\cp^{1/2}\pb;1)=f(\pa,\pb;1)$ so that finally
\begin{equation}
B_d=\smp^{d-2}\frac{dS_d2^{d-1}}{\cd^2}b_d\frac{N}{Z},
\end{equation}
where the constant $b_d$ is given by
\TCUT{
b_d\doteq
\int\frac{\d\pa\d\pb}{\pa^2-\pb^2}
|\pa+\pb|^{1-d}|\pa-\pb|^{1-d}
}{
f(\pa,\pb;1).
}
Remarkably, this integral appears in the second-order exchange contribution to the correlation energy of the homogeneous electron gas \cite{Gell-MannBrueckner-PR106}. First computed numerically, its value was then obtained in closed form for $d=3$ by Onsager \etal\ \cite{OnsagerMittagStephen-AnnPhysG473}, for $d=2$ by Isihara and Ioriatti \cite{IsiharaIoriatti-PRB22}, and for any $d$ by Glasser \cite{Glasser-JCAM10}. Effectively,
\begin{equation}
B_3=
\frac{\smp N}{Z}
\Big[\frac{1}{6}\ln 2-\frac{3}{4\pi^2}\zeta(3)\Big],
\end{equation}
and
\begin{equation}
B_2=
\frac{N}{Z}
\Big[\frac{G}{3}-\frac{2\gamma}{\pi^2}\Big],
\end{equation}
where $G$ is Catalan's constant and 
\begin{equation}
\gamma=
\sum_{n=0}^\infty\frac{(-1)^n}{(n+1)^3}
\sum_{m=0}^n\frac{(-1)^m}{2m+1}.
\end{equation}

\ITE{
\begin{widetext}
To compute $C_d$, we write $f\FX$ semiclassically as
\begin{multline}
-f\FX=
\frac{2}{\smp^{d+1}\cd^3}
\int\d\ka\d\kb\d\qv\d\zv\D\rv
g\big(\qv;\cp(\xv+\smp\zv)\big)
e^{i(\rv,\qv-\ka)-i(\zv,\kb)}
\\\times
\frac{
\hs{\big(\ka+\tfrac\kb2\big)^2-\cp(\xv+\smp(\tfrac\zv2+\tfrac\rv4))}
\hs{\cp(\xv+\smp(\tfrac\zv2-\tfrac\rv4))-\big(\ka-\tfrac\kb2\big)^2}
}{
2(\ka,\kb)
+\cp(\xv+\smp(\tfrac\zv2-\tfrac\rv4))
-\cp(\xv+\smp(\tfrac\zv2+\tfrac\rv4))
}.
\end{multline}
\end{widetext}
}{
To compute $C_d$, we write $f\FX$ semiclassically as
\begin{multline}
-f\FX=
\frac{2}{\smp^{d+1}\cd^3}
\int\d\ka\d\kb\d\qv\d\zv\D\rv
g\big(\qv;\cp(\xv+\smp\zv)\big)
e^{i(\rv,\qv-\ka)-i(\zv,\kb)}
\\\times
\frac{
\hs{\big(\ka+\tfrac\kb2\big)^2-\cp(\xv+\smp(\tfrac\zv2+\tfrac\rv4))}
\hs{\cp(\xv+\smp(\tfrac\zv2-\tfrac\rv4))-\big(\ka-\tfrac\kb2\big)^2}
}{
2(\ka,\kb)
+\cp(\xv+\smp(\tfrac\zv2-\tfrac\rv4))
-\cp(\xv+\smp(\tfrac\zv2+\tfrac\rv4))
}.
\end{multline}
}
The limit $\smp\to0$ in this expression gives
\begin{equation}
-f\FX=
\frac{1}{\smp^{d+1}\cd}
\inta\qv
g(\qv;\cp\FX)
\ks{\qv^2-\cp\FX},
\end{equation}
so that 
\TCUT{
C_d=
\smp^{d-2}\frac{4}{\cd^2}
\int
\frac{\d\xa\d\xb}{\norm{\xa-\xb}}
}{
\cp_+^{(d-1)/2}(\xa)
\cp_+^{(d-1)/2}(\xb).
}
It is a priori surprising that such a term of order $Z^{-3}$ gives semiclassically a result of the same order as those of order $Z^{-3}$. Indeed, all the other terms of order $Z^{-3}$ gave semiclassically a result of the order $Z^{-3}\smp^{d+3}$, up to $\ln\smp\inv$ corrections. We assume that all the term of higher order in $Z\inv$ give semiclassically a result of the order $Z^{-n}\smp^{d+n}$, up to corrections in $\ln\smp\inv$. But we have not proven it, and this question remains to be settled.

To summarize, in the case of atoms, and using the rotational symmetry of TF local chemical potential, we find
\OCUT{
-\ECB=
0.06390\ \frac{N}{Z^{4/3}}
}{
-\frac{1}{Z^{1/3}}\frac{2(2\pi)^3}{3}
\int_0^\infty\D{r}r\int_0^r\D{s}s^2\cp_+(r)\cp_+(s),
\label{ECB:f}
}
and in the case of dots, we find
\OCUT{
-\ECB=0.1455
}{
-\frac{1}{2\pi^4}
\int\frac{\d\xa\d\xb}{\norm{\xa-\xv}}
\cp_+^{1/2}(\xa)\cp_+^{1/2}(\xb).
}
%
%
%
%
\section{Correlation Energy of Neutral Atoms: Comparison With Numerical and Experimental Values}
Let us first recall that we have decomposed the ground state energy as $\sEQ=\sEHX+\sECR$. While $\sEHX$ looks like the HF energy $\sEHF$, the two energies differ in their inclusion of exchange effects, and shouldn't be confused. Taking a determinant of Hartree wave functions for the trial wave function, we will get $\sEHX$ for the energy. Therefore, we have the inequality $\sEHF\leq\sEHX$. As shown by a semiclassical HF theory, the difference $\sEHX-\sEHF$ is zero up to the order $N^{5/3}$ \cite{Englert-1988}. We have therefore shown that HF is correct up to order $N^{5/3}$, a result that has already been rigorously proven by Fefferman and Seco \cite{FeffermanSeco-HPA71,FeffermanSeco-AdvMath107a}. However, we do not know whether $\sEHX$ and $\sEHF$ are equivalent up to order $N$. Therefore, we do not know if $\ECR$ corresponds to what is commonly referred to as the correlation energy, that is the HF-relative energy $\smash{\sECRhf\doteq\sEQ-\sEHF}$, up to order $N$.

\begin{figure}[t]
\centering
\begin{small}\begin{psfrags}
\def\PFGstripminus-#1{#1}%
\def\PFGshift(#1,#2)#3{\raisebox{#2}[\height][\depth]{\hbox{%
  \ifdim#1<0pt\kern#1 #3\kern\PFGstripminus#1\else\kern#1 #3\kern-#1\fi}}}%
\providecommand{\PFGstyle}{}%
%
\psfrag{CellTextDa}[cc][cc]{\PFGstyle $-\bar{E}_\textsc{c}^\textsc{hf}/N$ (Ha)}%
\psfrag{ext}[cl][cl]{\PFGstyle \footnotesize{exp}}%
\psfrag{extHF}[cl][cl]{\PFGstyle \footnotesize{ext HF}}%
\psfrag{FormBoxNnT}[cc][cc]{\PFGstyle $N$}%
\psfrag{S0}[tc][tc]{\PFGstyle \footnotesize{$ 0$}}%
\psfrag{S10}[tc][tc]{\PFGstyle \footnotesize{$10$}}%
\psfrag{S20}[tc][tc]{\PFGstyle \footnotesize{$20$}}%
\psfrag{S30}[tc][tc]{\PFGstyle \footnotesize{$30$}}%
\psfrag{S40}[tc][tc]{\PFGstyle \footnotesize{$40$}}%
\psfrag{S50}[tc][tc]{\PFGstyle \footnotesize{$50$}}%
\psfrag{S60}[tc][tc]{\PFGstyle \footnotesize{$60$}}%
\psfrag{W000}[cr][cr]{\PFGstyle \footnotesize{$0.00$}}%
\psfrag{W001}[cr][cr]{\PFGstyle \footnotesize{$0.01$}}%
\psfrag{W002}[cr][cr]{\PFGstyle \footnotesize{$0.02$}}%
\psfrag{W003}[cr][cr]{\PFGstyle \footnotesize{$0.03$}}%
\psfrag{W004}[cr][cr]{\PFGstyle \footnotesize{$0.04$}}%
\psfrag{W005}[cr][cr]{\PFGstyle \footnotesize{$0.05$}}%
\psfrag{W006}[cr][cr]{\PFGstyle \footnotesize{$0.06$}}%
\psfrag{W007}[cr][cr]{\PFGstyle \footnotesize{$0.07$}}%
\includegraphics{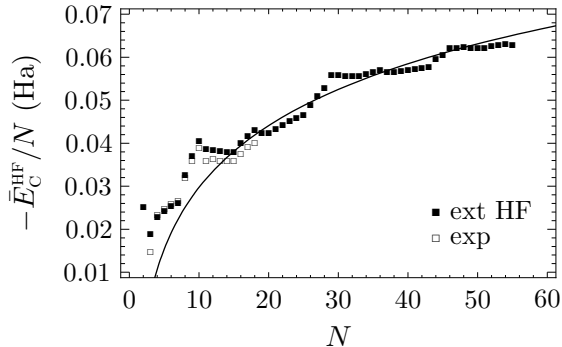}
\end{psfrags}\end{small}
\caption{Per electron HF-relative correlation energy for neutral atoms with up to $55$ electrons. Data points correspond to experimental values (exp) \cite{ChakravortyDavidson-JPC100} and extended HF values (ext HF) \cite{ClementiCorongiu-IJQC62} for $\sECRhf/N$. For sufficiently large atoms, the per electron energy $\sECRhf/N=-0.062\ \ln N-0.018$ (solid line) agrees with both experimental and extended HF values, presenting an essentially oscillating deviation of less than $8\%$ when $N\geq10$.}
\label{F1}
\end{figure}

Furthermore, while we have numerically computed the terms $x_a$ and $x_b$ of $\sECRa$ as well as the second term of the contribution $\sECRb$ in \eqref{ECB:f} for neutral atoms ---resulting in contributions to $-\sECR$, given in hartrees, of $0.06533\ N$, $-0.00329\ N$, and $-1.1044\ N$, respectively--- we haven't computed the constant $x$ appearing in the contribution $\sECRa$. Therefore, for neutral atoms, we have the correlation energy, given in hartrees, $\sECR=-0.062\ N\ln N+cN$, with $c$ to be determined. If we now assume that $\sEHF$ differs from $\sEHX$ at most by a contribution $\delta N$, we can compare the energy $\sECRhf=-0.062\ N\ln N+c'N$, with $c'=c+\delta$, to experimental and numerical values for $\sECRhf$. Experimental values exist for atoms containing up to $18$ electrons \cite{[{}] [{. The values for the HF-relative correlation energy are located in the supplemental pages, available online}]ChakravortyDavidson-JPC100}, and were obtained by removing from measured ground state energies the relativistic contribution; numerical values exist for atoms containing up to $55$ electrons \cite{ClementiCorongiu-IJQC62}, and were obtained in an extended HF approach. We see in figure \ref{F1}, that the $N\ln N$ term in $\sECR$ is essential in reproducing the behavior of reference values, and that with $c'=-0.018$, $\sECRhf$ agrees with experimental and numerical values, differing essentially by an oscillating contribution when $N$ is sufficiently large.
%
%
%
%
\section{Hartree-Exchange Energy of Quantum Dots}
Our remaining task is to compute the HX energy semiclassically, in the case of dots. It has been recognized \cite{BrackBhaduri-1997} that one should distinguish a smooth part and an oscillating part in a semiclassical expansion of the density matrix or the integrated density of states of a quantum system. The oscillating part, contrary to the smooth part, depends crucially on the nature of the classical dynamics associated to the potential. We will consider here only the smooth part, the oscillating one being discussed in another article.

The density matrix has the semiclassical expansion
\OCUT{
\nf\FF{\xv+\smp\rv/2}{\xv-\smp\rv/2}=
}{
\smp^{-2}\nfo(\rv;\cp\FX)+\nfi(\rv;\cp\FX)+\ord{\smp},
}
where
\begin{equation}
\nfo(\rv;\cp)\doteq
\frac{1}{(2\pi)^2}
\inta\pv e^{i\pv\rv}\hs{\cp-\pv^2}
\end{equation}
For $\nfi(\rv,\cp)$, we will only need here the fact that
\OCUT{
\inta\xv\nfi(0;\cp\FX)=
}{
-\frac{1}{48\pi}
\inta\xv\lap\W\FX\ks{\cp\FX}.
}
Consequently, we have, since $\idos(\cp)=\inta\xv\nf\FF\xv\xv$,
\OCUT{
\int^\cp\D{e}\idos(e)=
\frac{1}{8\pi\smp^2}
\inta\xv\cp_+^2\FX
}{
-\frac{1}{48\pi}
\inta\xv\lap\W\FX\hs{\cp\FX}.
}
This suggests the decomposition of $\cpo$ and $\W$, up to the order $\smp^2$,
\begin{align}
\cpo&=
\cpoo+\smp^2\cpoi,\\
\W\FX&=
\Wo\FX+\smp^2\Wi\FX,
\end{align}
and correspondingly
\begin{equation}
\deno\FX=
\denoo\FX+\smp^2\denoi\FX,
\end{equation}
with
\begin{equation}
\denoo\FX=
2\nfo(0;\cpoo\FX),
\end{equation}
and
\OCUT{
\denoi\FX=
2\nfi(0;\cpoo\FX)
}{
+2\frac{d}{d\cp}\nfo(0;\cp)\bigg|_{\cp=\cpoo\FX}
\cpoi\FX,
}
where $\cpoo\FX=\cpoo-\Wo\FX$ and $\cpoi\FX=\cpoi-\Wi\FX$. $\cpoo$ will be fixed by the constraint
\begin{equation}
1=
\inta\xv\denoo\FX,
\end{equation}
and $\cpoi$ by the constraint
\begin{equation}
0=
\inta\xv\denoi\FX.
\end{equation}
The self-consistent equation decomposes into
\begin{equation}
\Wo\FX=
\V\FX+\int\frac{\d\yv}{\norm{\xv-\yv}}\denoo\FY,
\end{equation}
and
\begin{equation}
\Wi\FX=
\int\frac{\d\yv}{\norm{\xv-\yv}}\denoi\FY.
\end{equation}
We can now decompose the Hartree part $\EH$ of the ground state energy using the fact that $Z=N$
\begin{equation}
\EH=
\Eoo+\smp^2\Eoi,
\end{equation}
where
\OCUT{
\Eoo=
\cpoo N-
\frac{N}{4\pi}
\inta\xv\cpoo^2\FX\hs{\cpoo\FX}
}{
-\frac{N}{2}\inta\xv\denoo\FX
\big(\Wo\FX-\V\FX\big),
}
and
\begin{multline}
\Eoi=
N\Big[
\cpoi-
\inta\xv\denoo\FX\cpoi\FX-
\inta\xv\denoo\FX\Wi\FX\Big]
\\
\shoveright{
+\frac{N}{24\pi}\inta\xv\lap\Wo\FX\hs{\cpoo\FX}}
\\
\shoveleft{
\phantom{\Eoi}=\frac{N}{24\pi}\inta\xv\lap\Wo\FX\hs{\cpoo\FX}.\hfill}
\end{multline}
The exchange term becomes
\OCUT{
\EXC=
-\frac{1}{Z\smp^3}
\int\frac{\d\rv}{r}
\inta\xv
\Big[\nfo(\rv;\cpoo\FX)
}{
+\smp^2\nfi(\rv;\cpoo\FX)\Big]^2,
}
or
\begin{equation}
\EXC=
-\frac{1}{\smp}
\frac{J}{(2\pi)^3}
\inta\xv\cpoop^{3/2}\FX
+\ord{\smp},
\end{equation}
where
\begin{equation}
J=
\int\frac{\d{\pv_1}\d{\pv_2}}{\norm{\pv_1+\pv_2}}
\hs{1-\pv_1^2}\hs{1-\pv_2^2}
=\frac{16\pi}{3}.
\end{equation}
Finally, the correction
\begin{equation}
\ECP=
\half
\frac{\big[\partial_\cp\EXC\big]^2}
{\PP_0''(\cp)}\bigg|_{\cp=\cpoo}
\end{equation}
gives
\begin{equation}
\ECP=
\frac{1}{\pi^3}
\frac{\displaystyle
\Big[\inta\xv\cpoo^{1/2}\FX a\FX\Big]^2
}{\displaystyle
\inta\xv a\FX\hs{\cpoo\FX}},
\end{equation}
where $a\FX$ is the solution of the linear integral equation
\begin{equation}
a\FX=
1-\frac{1}{2\pi}\int\frac{\d\yv}{\norm{\xv-\yv}}
a\FY\hs{\cp\FY}.
\label{LIEa}
\end{equation}
We can now summarize the results for the smooth part of the ground state energy corresponding to HX. Expressed for the original problem, we have
\ITE{
\begin{multline}
\sEHX=
N^2\ETF
+N^{3/2}\frac{2}{3\pi^2}\inta\xv\cp_+^{3/2}\FX
\\
+\frac{N}{24\pi}\inta\xv\lap\W\FX\hs{\cp\FX}\\
+\frac{N}{\pi^3}
\frac{\displaystyle
\Big[\inta\xv\cp_+^{1/2}\FX a\FX\Big]^2
}{\displaystyle
\inta\xv a\FX\hs{\cp\FX}},
\label{EHX:Q:f}
\end{multline}
}{
\begin{multline}
\sEHX=
N^2\ETF
+N^{3/2}\frac{2}{3\pi^2}\inta\xv\cp_+^{3/2}\FX
\\
+\frac{N}{24\pi}\inta\xv\lap\W\FX\hs{\cp\FX}
+\frac{N}{\pi^3}
\frac{\displaystyle
\Big[\inta\xv\cp_+^{1/2}\FX a\FX\Big]^2
}{\displaystyle
\inta\xv a\FX\hs{\cp\FX}},
\label{EHX:Q:f}
\end{multline}
}
where $\ETF$ is the TF energy
\begin{equation}
\ETF=
\half\Big[\cp
+\frac{1}{2\pi}\inta\xv\V\FX\cp_+\FX\Big],
\end{equation}
and $\cp$ and $\W\FX$ being solution of the TF equations
\begin{equation}
\W\FX=
\V\FX+\frac{1}{2\pi}\int\frac{\d\yv}{\norm{\xv-\yv}}\cp_+\FY,
\end{equation}
and
\begin{equation}
1=
\frac{1}{2\pi}
\inta\xv\cp_+\FX.
\end{equation}
%
%
%
%
\section{Hartree-Exchange Energy of Atoms}
In the case of atoms, it is needed to evaluate semiclassically the HX energy to the same order as the correlation energy. One can take advantage of the spherical symmetry of the potential $\W\FX$, thus reducing the problem to a one-dimensional one, which can be studied semiclassically by WKB type techniques. However, a specific difficulty arises in the atomic case. The potential $\W(r)$ diverges like the Coulomb potential $r\inv$ near the origin. In physical terms, the semiclassical limit has to be reconsidered for strongly bound electrons. This is at the origin of the Scott and Schwinger corrections, which have successfully handled this problem. But one needs more, namely to compute the integrated density of states up to the order $\smp$. Moreover, the exchange energy, whose dominant term has been computed by Dirac \cite{Dirac-PCPS26} requires a knowledge of the density matrix $\nf\FXY$ up to the order $\smp^2$. If we use a standard expression for this correction of the density matrix, we get a logarithmically divergent correction for the exchange energy. The origin of this divergence is the slow decay of the Coulomb potential appearing in the expression of the exchange energy.

As in the case of dots, the energy will be decomposed into an oscillating and a smooth part, $\sEHX=
\sEHXsm+\sEHXos$. We expect for $\sEHXsm$ a neutral atom an asymptotic expansion given by
\begin{equation}
\sEHXsm=
\sum_{j=3}^7c_jN^j+c_0N\ln N,
\label{EHXsA}
\end{equation}
with the constants, expressed in hartrees,
\begin{equation}
c_7=-0.7687,\quad c_6=-0.5,\quad c_5=-0.2699,
\end{equation}
known \cite{Morgan-1996}. We have not undertaken the task of computing $c_4$, $c_3$, and $c_0$. 

The oscillating part, of order $N^{4/3}$, has been computed by Schwinger and Englert \cite{EnglertSchwinger-PRA32c}. In the case of atoms we see therefore, that contrary to the case of dots, it is more important than the correlation energy. This oscillating part can be understood as the first appearance of shell effects, in an atom described as a liquid by the smooth part of the energy. This is the interpretation of Schwinger and Englert. But it also has a dynamical interpretation. Indeed it is standard now to decompose the density of states $\dos(e)=\partial_e\idos(e)$ into two parts semiclassically \cite{BrackBhaduri-1997} as
\begin{equation}
\dos(e)=
\dossm(e)+
\dosos(e).
\end{equation}
The smooth part $\dossm(e)$ is given by an asymptotic expansion in $\smp\inv$, whose coefficients are some integrals depending on $\W\FX$. The corresponding part of $\sEHX$ was given in \eqref{EHXsA}. The oscillating part
\begin{equation}
\dosos(e)=
\sum_{\gamma}A_\smp(e,\gamma)\cos\Big(\frac{1}{\smp}S(e,\gamma)+\sigma_\gamma\frac{\pi}{2}\Big)
\end{equation}
is given by a sum over the periodic orbits $\gamma$ of a classical particle moving in the potential $\W\FX$, where $S(e,\gamma)$ is the classical action along the orbit, $\sigma_\gamma$ is the orbit's Maslov index, and $A_\smp(e,\gamma)$ depends on the orbit's period and stability. $\dosos(e)$ therefore depends crucially on the nature of the dynamics associated to $\W\FX$. In the case of atoms, the TF potential $\W\FX$ is rotationally symmetric, so that the dynamics is that of an integrable system. In this case the periodic orbits are stable and degenerate. A general formula for $\dosos(e)$, in the case of a rotationally symmetric potential, is given in \cite{BerryMount-RPP35}. One can see that it corresponds to the formula obtained by Schwinger and Englert. In fact, all these authors base their analysis on the use of the Poisson formula
\begin{equation}
\sum_{n\in\mathbb{Z}}\ks{x-n}=
\sum_{m\in\mathbb{Z}} e^{i2\pi mx}
\end{equation}
%
%
%
%
\section{Conclusion}
We have provided a method to systematically compute the ground state energy of non-relativistic atoms and quantum dots, by means of an asymptotic expansion in $\smash{N\inv}$, whose terms have have to be evaluated semiclassically. The dominant terms are give by a semiclassical HX theory, which coincides with HF theory only up to a certain order in $\smash{N\inv}$. Correlation effects go beyond HF. They are of order $N\ln N$ for atoms, and order $N$ for dots. It remains in the case of atoms to compute numerically a constant appearing in the term of order $N$ in the correlation energy. In the case of atoms, it remains also to fully compute the HX energy up to the order $N$. This represents a challenge in semiclassical physics. This computation would allow a better comparison with the data, because when we used our results for the correlation energy, this quantity was defined as the difference between the true energy and the HF energy, and we have seen that it is not the full HF energy which matters, but rather the HX energy.

Among the possible extensions of this work in the case of atoms are the following: (a) For non-relativistic atoms, and with $\bm{L}$ the total angular momentum and $\bm{S}$ the spin, $\bm{L}^2$, $L_z$, $\bm{S}^2$, and $S_z$ are conserved. It would therefore be interesting to compute the ground state energy with these quantities being fixed; (b) Compute the ground state energy taking into account the dominant relativistic corrections, which become more important when $N$ is large.

Experimental results for quantum dots in the presence of a magnetic field have been obtained \cite{TaruchaEA-PRL77}. A TF type theory has already been established in this case \cite{LiebSolovejYngvason-PRB51}. It is therefore an interesting, though challenging, problem to extend our results for the corrections to TF theory to this situation.
%
%
%
%
\section{Acknowledgements}
We thank C.~Plocek for useful and stimulating discussions at the early stage of this work. This work was supported by the Fonds National Suisse de la Recherche Scientifique.
%
%
%
%
\section{Appendix}
In the case of atoms, the computation of $\ECS$ is rather delicate, so that we give here a summary of the main steps in this computation. The crucial quantity to compute is what we called $X_\smp$, and given by
\TCUT{
X_\smp=
4\intb\kv\pv
\big(\kv+\smp\tfrac{\pv}{2}\big)^{-2}
\big(\kv-\smp\tfrac{\pv}{2}\big)^{-2}
}{
\intin\D{t}\norm{\Gamma_t\FF\kv\pv}^2
}
with
\begin{equation}
\Gamma_t\FF\kv\pv=
\inta\xv e^{i\pv\xv}
\alpha_t(\kv;\cp\FX),
\end{equation}
where we recall that
\TCUT{
\alpha_t(\kv;\cp)=
\inta\qv
e^{-t[(\qv+\kv)^2-\qv^2]}
}{
\hs{(\kv+\qv)^2-\cp}
\hs{\cp-\qv^2}.
}
$\Gamma_t\FF\kv\pv$ depends only on $k$ and $\pv$, so that
\begin{equation}
X_\smp=
8\pi\inta\pv Y_\smp(\pv),
\end{equation}
where
\begin{equation}
Y_\smp(\pv)\doteq
\intin\d{k}
\frac{k^2}{k^2+1}
\ln\bigg|\frac{k+1}{k-1}\bigg|
\gamma\big(\smp\tfrac{kp}{2}\big|\pv\big),
\end{equation}
where
\begin{equation}
\gamma\FF k\pv
\intin\d{t}
\frac{\norm{\Gamma_t\FF\kv\pv}^2}{k}.
\end{equation}
We then decompose $Y_\smp$ into four parts. Let
\ITE{
\begin{align}
Y_{1;\smp}(\pv)&\doteq
\intin\d{k}
\bigg[
\frac{k^2}{k^2+1}
\ln\bigg|\frac{k+1}{k-1}\bigg|
\nonumber\\
&\qquad\qquad-\hs{k-1}\frac{2}{k}\bigg]
\gamma\big(\smp\tfrac{kp}{2}\big|\pv\big)\\
Y_{2;\smp}(\pv)&\doteq
2\int_1^\infty\frac{\d{k}}{k}
\gamma\big(\smp\tfrac{kp}{2}\big|\pv\big)\\
Y_{3;\smp}(\pv)&\doteq
2\int_{\smp p/2}^1\frac{\d{k}}{k}
\Big[
\gamma\big(\smp\tfrac{kp}{2}\big|\pv\big)-
\gamma\big(0\big|\pv\big)
\Big]\\
Y_{4;\smp}(\pv)&\doteq
2\int_{\smp p/2}^1\frac{\d{k}}{k}
\gamma\big(0\big|\pv\big).
\end{align}
}{
\begin{align}
Y_{1;\smp}(\pv)&\doteq
\intin\d{k}
\bigg[
\frac{k^2}{k^2+1}
\ln\bigg|\frac{k+1}{k-1}\bigg|
-\hs{k-1}\frac{2}{k}\bigg]
\gamma\big(\smp\tfrac{kp}{2}\big|\pv\big)\\
Y_{2;\smp}(\pv)&\doteq
2\int_1^\infty\frac{\d{k}}{k}
\gamma\big(\smp\tfrac{kp}{2}\big|\pv\big)\\
Y_{3;\smp}(\pv)&\doteq
2\int_{\smp p/2}^1\frac{\d{k}}{k}
\Big[
\gamma\big(\smp\tfrac{kp}{2}\big|\pv\big)-
\gamma\big(0\big|\pv\big)
\Big]\\
Y_{4;\smp}(\pv)&\doteq
2\int_{\smp p/2}^1\frac{\d{k}}{k}
\gamma\big(0\big|\pv\big).
\end{align}
}
This decomposition is justified by the fact that it can be shown that $\gamma\FF0\pv$ is finite and $\gamma\FF{k}\pv=\ord{k^{-3}}$. For these proofs one needs to remember that $\cp(r)\sim r\inv$ when $r\to0$ and $\cp(r)\sim r^{-4}$ when $r\to\infty$. The decomposition allows us to get the asymptotic behavior of $X_\smp$
\begin{equation}
X_\smp=A\ln\smp\inv+B,
\end{equation}
where
\begin{equation}
A=16\pi\inta\pv\gamma\FF0\pv,
\end{equation}
and $B=\sum_{j=1}^4B_j$, with
\begin{align}
B_1&\doteq
8\pi C
\inta\pv\gamma\FF0\pv\\
B_2&\doteq
16\pi\inta\pv
\int_1^\infty\frac{\d{k}}{k}
\gamma\FF{k}\pv\\
B_3&\doteq
16\pi\inta\pv
\intoi\frac{\d{k}}{k}
\big[\gamma\FF{k}\pv-\gamma\FF0\pv\big]\\
B_4&\doteq
4\pi\inta\pv\ln(2/p)
\gamma\big(0\big|\pv\big),
\end{align}
and
\begin{equation}
C=
\intin\d{k}\bigg[
\frac{k^2}{k^2+1}
\ln\bigg|\frac{k+1}{k-1}\bigg|-
\frac2k\hs{k-1}\bigg].
\end{equation}
other work is needed, that we do not reproduce here but can be found in \cite{Rueedi-2009}, to express these constants in terms of $\cp(r)$, then obtain expression \eqref{X0} by using the normalization condition \eqref{cp:normal} where applicable.
%
%
%
%
%

\begin{thebibliography}{10}%
\makeatletter
\providecommand \@ifxundefined [1]{%
 \ifx #1\undefined \expandafter \@firstoftwo
 \else \expandafter \@secondoftwo
\fi
}%
\providecommand \@ifnum [1]{%
 \ifnum #1\expandafter \@firstoftwo
 \else \expandafter \@secondoftwo
\fi
}%
\providecommand \enquote [1]{``#1''}%
\providecommand \bibnamefont  [1]{#1}%
\providecommand \bibfnamefont [1]{#1}%
\providecommand \citenamefont [1]{#1}%
\providecommand\href[0]{\@sanitize\@href}%
\providecommand\@href[1]{\endgroup\@@startlink{#1}\endgroup\@@href}%
\providecommand\@@href[1]{#1\@@endlink}%
\providecommand \@sanitize [0]{\begingroup\catcode`\&12\catcode`\#12\relax}%
\@ifxundefined \pdfoutput {\@firstoftwo}{%
 \@ifnum{\z@=\pdfoutput}{\@firstoftwo}{\@secondoftwo}%
}{%
 \providecommand\@@startlink[1]{\leavevmode}%
 \providecommand\@@endlink[0]{}%
}{%
 \providecommand\@@startlink[1]{%
  \leavevmode
  \pdfstartlink
   attr{/Border[0 0 1 ]/H/I/C[0 1 1]}%
   user{/Subtype/Link/A<</Type/Action/S/URI/URI(#1)>>}%
  \relax
 }%
 \providecommand\@@endlink[0]{\pdfendlink}%
}%
\providecommand \url  [0]{\begingroup\@sanitize \@url }%
\providecommand \@url [1]{\endgroup\@href {#1}{\urlprefix}}%
\providecommand \urlprefix [0]{URL }%
\providecommand \Eprint[0]{\href }%
\@ifxundefined \urlstyle {%
  \providecommand \doi [1]{doi:\discretionary{}{}{}#1}%
}{%
  \providecommand \doi [0]{doi:\discretionary{}{}{}\begingroup
  \urlstyle{rm}\Url }%
}%
\providecommand \doibase [0]{http://dx.doi.org/}%
\providecommand \Doi[1]{\href{\doibase#1}}%
\providecommand \bibAnnote [3]{%
  \BibitemShut{#1}%
  \begin{quotation}\noindent
    \textsc{Key:}\ #2\\\textsc{Annotation:}\ #3%
  \end{quotation}%
}%
\providecommand \bibAnnoteFile [2]{%
  \IfFileExists{#2}{\bibAnnote {#1} {#2} {\input{#2}}}{}%
}%
\providecommand \typeout [0]{\immediate \write \m@ne }%
\providecommand \selectlanguage [0]{\@gobble}%
\providecommand \bibinfo [0]{\@secondoftwo}%
\providecommand \bibfield [0]{\@secondoftwo}%
\providecommand \translation [1]{[#1]}%
\providecommand \BibitemOpen[0]{}%
\providecommand \bibitemStop [0]{}%
\providecommand \bibitemNoStop [0]{.\EOS\space}%
\providecommand \EOS [0]{\spacefactor3000\relax}%
\providecommand \BibitemShut [1]{\csname bibitem#1\endcsname}%
\bibitem{Thomas-PCPS23}%
  \BibitemOpen
  \bibfield{author}{%
  \bibinfo {author} {\bibfnamefont{L.~H.}\ \bibnamefont{Thomas}},\ }%
  \bibfield{journal}{%
  \bibinfo {journal} {Proc. Camb. Phil. Soc.}\ }%
  \textbf{\bibinfo {volume} {23}},\ \bibinfo {pages} {542} (\bibinfo {year}
  {1927})%
  \bibAnnoteFile{NoStop}{Thomas-PCPS23}%
\bibitem{Fermi-RANL6}%
  \BibitemOpen
  \bibfield{author}{%
  \bibinfo {author} {\bibfnamefont{E.}~\bibnamefont{Fermi}},\ }%
  \bibfield{journal}{%
  \bibinfo {journal} {Rend. Accad. Naz. Lincei}\ }%
  \textbf{\bibinfo {volume} {6}},\ \bibinfo {pages} {602} (\bibinfo {year}
  {1927})%
  \bibAnnoteFile{NoStop}{Fermi-RANL6}%
\bibitem{Morgan-1996}%
  \BibitemOpen
  \bibfield{author}{%
  \bibinfo {author} {\bibfnamefont{J.~D.}\ \bibnamefont{Morgan}},\ }%
  in\ \emph{\bibinfo {booktitle} {Atomic, Molecular, \& Optical Physics
  Handbook}},\ \bibinfo {editor} {edited by\ \bibinfo {editor}
  {\bibfnamefont{G.~W.~F.}\ \bibnamefont{Drake}}}\ (\bibinfo {publisher} {AIP
  Press},\ \bibinfo {year} {1996})\ Chap.~\bibinfo {chapter} {10}, pp.\
  \bibinfo {pages} {233--242}%
  \bibAnnoteFile{NoStop}{Morgan-1996}%
\bibitem{Dirac-PCPS26}%
  \BibitemOpen
  \bibfield{author}{%
  \bibinfo {author} {\bibfnamefont{P.~A.~M.}\ \bibnamefont{Dirac}},\ }%
  \bibfield{journal}{%
  \bibinfo {journal} {Proc. Camb. Philos. Soc.}\ }%
  \textbf{\bibinfo {volume} {26}},\ \bibinfo {pages} {376} (\bibinfo {year}
  {1930})%
  \bibAnnoteFile{NoStop}{Dirac-PCPS26}%
\bibitem{Scott-PM43}%
  \BibitemOpen
  \bibfield{author}{%
  \bibinfo {author} {\bibfnamefont{J.~M.~C.}\ \bibnamefont{Scott}},\ }%
  \bibfield{journal}{%
  \bibinfo {journal} {Philo. Mag.}\ }%
  \textbf{\bibinfo {volume} {43}},\ \bibinfo {pages} {859} (\bibinfo {year}
  {1952})%
  \bibAnnoteFile{NoStop}{Scott-PM43}%
\bibitem{Schwinger-PRA22}%
  \BibitemOpen
  \bibfield{author}{%
  \bibinfo {author} {\bibfnamefont{J.}~\bibnamefont{Schwinger}},\ }%
  \bibfield{journal}{%
  \Doi{10.1103/PhysRevA.22.1827}{\bibinfo {journal} {Phys. Rev. A}}\ }%
  \textbf{\bibinfo {volume} {22}},\ \bibinfo {pages} {1827} (\bibinfo {year}
  {1980})%
  \bibAnnoteFile{NoStop}{Schwinger-PRA22}%
\bibitem{Schwinger-PRA24}%
  \BibitemOpen
  \bibfield{author}{%
  \bibinfo {author} {\bibfnamefont{J.}~\bibnamefont{Schwinger}},\ }%
  \bibfield{journal}{%
  \Doi{10.1103/PhysRevA.24.2353}{\bibinfo {journal} {Phys. Rev. A}}\ }%
  \textbf{\bibinfo {volume} {24}},\ \bibinfo {pages} {2353} (\bibinfo {year}
  {1981})%
  \bibAnnoteFile{NoStop}{Schwinger-PRA24}%
\bibitem{EnglertSchwinger-PRA26}%
  \BibitemOpen
  \bibfield{author}{%
  \bibinfo {author} {\bibfnamefont{B.-G.}\ \bibnamefont{Englert}}\ and\
  \bibinfo {author} {\bibfnamefont{J.}~\bibnamefont{Schwinger}},\ }%
  \bibfield{journal}{%
  \Doi{10.1103/PhysRevA.26.2322}{\bibinfo {journal} {Phys. Rev. A}}\ }%
  \textbf{\bibinfo {volume} {26}},\ \bibinfo {pages} {2322} (\bibinfo {year}
  {1982})%
  \bibAnnoteFile{NoStop}{EnglertSchwinger-PRA26}%
\bibitem{EnglertSchwinger-PRA29a}%
  \BibitemOpen
  \bibfield{author}{%
  \bibinfo {author} {\bibfnamefont{B.-G.}\ \bibnamefont{Englert}}\ and\
  \bibinfo {author} {\bibfnamefont{J.}~\bibnamefont{Schwinger}},\ }%
  \bibfield{journal}{%
  \Doi{10.1103/PhysRevA.29.2331}{\bibinfo {journal} {Phys. Rev. A}}\ }%
  \textbf{\bibinfo {volume} {29}},\ \bibinfo {pages} {2331} (\bibinfo {year}
  {1984})%
  \bibAnnoteFile{NoStop}{EnglertSchwinger-PRA29a}%
\bibitem{EnglertSchwinger-PRA29b}%
  \BibitemOpen
  \bibfield{author}{%
  \bibinfo {author} {\bibfnamefont{B.-G.}\ \bibnamefont{Englert}}\ and\
  \bibinfo {author} {\bibfnamefont{J.}~\bibnamefont{Schwinger}},\ }%
  \bibfield{journal}{%
  \Doi{10.1103/PhysRevA.29.2339}{\bibinfo {journal} {Phys. Rev. A}}\ }%
  \textbf{\bibinfo {volume} {29}},\ \bibinfo {pages} {2339} (\bibinfo {year}
  {1984})%
  \bibAnnoteFile{NoStop}{EnglertSchwinger-PRA29b}%
\bibitem{EnglertSchwinger-PRA29c}%
  \BibitemOpen
  \bibfield{author}{%
  \bibinfo {author} {\bibfnamefont{B.-G.}\ \bibnamefont{Englert}}\ and\
  \bibinfo {author} {\bibfnamefont{J.}~\bibnamefont{Schwinger}},\ }%
  \bibfield{journal}{%
  \Doi{10.1103/PhysRevA.29.2353}{\bibinfo {journal} {Phys. Rev. A}}\ }%
  \textbf{\bibinfo {volume} {29}},\ \bibinfo {pages} {2353} (\bibinfo {year}
  {1984})%
  \bibAnnoteFile{NoStop}{EnglertSchwinger-PRA29c}%
\bibitem{LiebSimon-PRL31}%
  \BibitemOpen
  \bibfield{author}{%
  \bibinfo {author} {\bibfnamefont{E.~H.}\ \bibnamefont{Lieb}}\ and\ \bibinfo
  {author} {\bibfnamefont{B.}~\bibnamefont{Simon}},\ }%
  \bibfield{journal}{%
  \Doi{10.1103/PhysRevLett.31.681}{\bibinfo {journal} {Phys. Rev. Lett.}}\ }%
  \textbf{\bibinfo {volume} {31}},\ \bibinfo {pages} {681} (\bibinfo {year}
  {1973})%
  \bibAnnoteFile{NoStop}{LiebSimon-PRL31}%
\bibitem{LiebSimon-AdvMath23}%
  \BibitemOpen
  \bibfield{author}{%
  \bibinfo {author} {\bibfnamefont{E.~H.}\ \bibnamefont{Lieb}}\ and\ \bibinfo
  {author} {\bibfnamefont{B.}~\bibnamefont{Simon}},\ }%
  \bibfield{journal}{%
  \Doi{10.1016/0001-8708(77)90108-6}{\bibinfo {journal} {Adv. in Math.}}\ }%
  \textbf{\bibinfo {volume} {23}},\ \bibinfo {pages} {22} (\bibinfo {year}
  {1977})%
  \bibAnnoteFile{NoStop}{LiebSimon-AdvMath23}%
\bibitem{FeffermanSeco-AdvMath107a}%
  \BibitemOpen
  \bibfield{author}{%
  \bibinfo {author} {\bibfnamefont{C.~L.}\ \bibnamefont{Fefferman}}\ and\
  \bibinfo {author} {\bibfnamefont{L.~A.}\ \bibnamefont{Seco}},\ }%
  \bibfield{journal}{%
  \Doi{10.1006/aima.1994.1060}{\bibinfo {journal} {Adv. in Math.}}\ }%
  \textbf{\bibinfo {volume} {107}},\ \bibinfo {pages} {1} (\bibinfo {year}
  {1994})%
  \bibAnnoteFile{NoStop}{FeffermanSeco-AdvMath107a}%
\bibitem{FeffermanSeco-HPA71}%
  \BibitemOpen
  \bibfield{author}{%
  \bibinfo {author} {\bibfnamefont{C.~L.}\ \bibnamefont{Fefferman}}\ and\
  \bibinfo {author} {\bibfnamefont{L.~A.}\ \bibnamefont{Seco}},\ }%
  \bibfield{journal}{%
  \bibinfo {journal} {Helv. Phys. Act.}\ }%
  \textbf{\bibinfo {volume} {71}},\ \bibinfo {pages} {1} (\bibinfo {year}
  {1997})%
  \bibAnnoteFile{NoStop}{FeffermanSeco-HPA71}%
\bibitem{EnglertSchwinger-PRA32c}%
  \BibitemOpen
  \bibfield{author}{%
  \bibinfo {author} {\bibfnamefont{B.-G.}\ \bibnamefont{Englert}}\ and\
  \bibinfo {author} {\bibfnamefont{J.}~\bibnamefont{Schwinger}},\ }%
  \bibfield{journal}{%
  \Doi{10.1103/PhysRevA.32.47}{\bibinfo {journal} {Phys. Rev. A}}\ }%
  \textbf{\bibinfo {volume} {32}},\ \bibinfo {pages} {47} (\bibinfo {year}
  {1985})%
  \bibAnnoteFile{NoStop}{EnglertSchwinger-PRA32c}%
\bibitem{KouwenhovenAustingTarucha-RPP64}%
  \BibitemOpen
  \bibfield{author}{%
  \bibinfo {author} {\bibfnamefont{L.~P.}\ \bibnamefont{Kouwenhoven}}, \bibinfo
  {author} {\bibfnamefont{D.~G.}\ \bibnamefont{Austing}},\ and\ \bibinfo
  {author} {\bibfnamefont{S.}~\bibnamefont{Tarucha}},\ }%
  \bibfield{journal}{%
  \Doi{10.1088/0034-4885/64/6/201}{\bibinfo {journal} {Rep. Prog. Phys.}}\ }%
  \textbf{\bibinfo {volume} {64}},\ \bibinfo {pages} {701} (\bibinfo {year}
  {2001})%
  \bibAnnoteFile{NoStop}{KouwenhovenAustingTarucha-RPP64}%
\bibitem{ReimannManninen-RMP74}%
  \BibitemOpen
  \bibfield{author}{%
  \bibinfo {author} {\bibfnamefont{S.~M.}\ \bibnamefont{Reimann}}\ and\
  \bibinfo {author} {\bibfnamefont{M.}~\bibnamefont{Manninen}},\ }%
  \bibfield{journal}{%
  \Doi{10.1103/RevModPhys.74.1283}{\bibinfo {journal} {Rev. Mod. Phys.}}\ }%
  \textbf{\bibinfo {volume} {74}},\ \bibinfo {pages} {1283} (\bibinfo {year}
  {2002})%
  \bibAnnoteFile{NoStop}{ReimannManninen-RMP74}%
\bibitem{LiebSolovejYngvason-PRB51}%
  \BibitemOpen
  \bibfield{author}{%
  \bibinfo {author} {\bibfnamefont{E.~H.}\ \bibnamefont{Lieb}}, \bibinfo
  {author} {\bibfnamefont{J.~P.}\ \bibnamefont{Solovej}},\ and\ \bibinfo
  {author} {\bibfnamefont{J.}~\bibnamefont{Yngvason}},\ }%
  \bibfield{journal}{%
  \Doi{10.1103/PhysRevB.51.10646}{\bibinfo {journal} {Phys. Rev. B}}\ }%
  \textbf{\bibinfo {volume} {51}},\ \bibinfo {pages} {10646} (\bibinfo {year}
  {1995})%
  \bibAnnoteFile{NoStop}{LiebSolovejYngvason-PRB51}%
\bibitem{LebowitzPenrose-JMP7}%
  \BibitemOpen
  \bibfield{author}{%
  \bibinfo {author} {\bibfnamefont{J.~L.}\ \bibnamefont{Lebowitz}}\ and\
  \bibinfo {author} {\bibfnamefont{O.}~\bibnamefont{Penrose}},\ }%
  \bibfield{journal}{%
  \Doi{10.1063/1.1704821}{\bibinfo {journal} {J. Math. Phys.}}\ }%
  \textbf{\bibinfo {volume} {7}},\ \bibinfo {pages} {98} (\bibinfo {year}
  {1966})%
  \bibAnnoteFile{NoStop}{LebowitzPenrose-JMP7}%
\bibitem{Lieb-JMP7}%
  \BibitemOpen
  \bibfield{author}{%
  \bibinfo {author} {\bibfnamefont{E.~H.}\ \bibnamefont{Lieb}},\ }%
  \bibfield{journal}{%
  \Doi{doi:10.1063/1.1704992}{\bibinfo {journal} {J. Math. Phys.}}\ }%
  \textbf{\bibinfo {volume} {7}},\ \bibinfo {pages} {1016} (\bibinfo {year}
  {1966})%
  \bibAnnoteFile{NoStop}{Lieb-JMP7}%
\bibitem{Englert-1988}%
  \BibitemOpen
  \bibfield{author}{%
  \bibinfo {author} {\bibfnamefont{B.-G.}\ \bibnamefont{Englert}},\ }%
  \emph{\bibinfo {title} {{S}emiclassical {T}heory of {A}toms}},\ Lecture Notes
  in Physics\ (\bibinfo {publisher} {Springer-Verlag},\ \bibinfo {year}
  {1988})%
  \bibAnnoteFile{NoStop}{Englert-1988}%
\bibitem{ClementiCorongiu-IJQC62}%
  \BibitemOpen
  \bibfield{author}{%
  \bibinfo {author} {\bibfnamefont{E.}~\bibnamefont{Clementi}}\ and\ \bibinfo
  {author} {\bibfnamefont{G.}~\bibnamefont{Corongiu}},\ }%
  \bibfield{journal}{%
  \Doi{10.1002/(SICI)1097-461X(1997)62:6<571::AID-QUA2>3.0.CO;2-T}{\bibinfo
  {journal} {Int. J. Quantum Chem.}}\ }%
  \textbf{\bibinfo {volume} {62}},\ \bibinfo {pages} {571} (\bibinfo {year}
  {1997})%
  \bibAnnoteFile{NoStop}{ClementiCorongiu-IJQC62}%
\bibitem{NegeleOrland-1988}%
  \BibitemOpen
  \bibfield{author}{%
  \bibinfo {author} {\bibfnamefont{J.~W.}\ \bibnamefont{Negele}}\ and\ \bibinfo
  {author} {\bibfnamefont{H.}~\bibnamefont{Orland}},\ }%
  \emph{\bibinfo {title} {{Q}uantum {M}any-{P}article {S}ystems}},\ Frontiers
  in Physics\ (\bibinfo {publisher} {Addison-Wesley Publishing Company},\
  \bibinfo {year} {1988})%
  \bibAnnoteFile{NoStop}{NegeleOrland-1988}%
\bibitem{Rueedi-2009}%
  \BibitemOpen
  \bibfield{author}{%
  \bibinfo {author} {\bibfnamefont{R.}~\bibnamefont{Rueedi}},\ }%
  \emph{\bibinfo {title} {{G}round {S}tate {P}roperties of {L}arge {A}toms and
  {Q}uantum {D}ots}},\ Ph.D. thesis,\ \bibinfo {school} {Ecole Polytechnique
  F\'ed\'erale de Lausanne} (\bibinfo {year} {2009})%
  \bibAnnoteFile{NoStop}{Rueedi-2009}%
\bibitem{Gell-MannBrueckner-PR106}%
  \BibitemOpen
  \bibfield{author}{%
  \bibinfo {author} {\bibfnamefont{M.}~\bibnamefont{Gell-Mann}}\ and\ \bibinfo
  {author} {\bibfnamefont{K.~A.}\ \bibnamefont{Brueckner}},\ }%
  \bibfield{journal}{%
  \Doi{10.1103/PhysRev.106.364}{\bibinfo {journal} {Phys. Rev.}}\ }%
  \textbf{\bibinfo {volume} {106}},\ \bibinfo {pages} {364} (\bibinfo {year}
  {1957})%
  \bibAnnoteFile{NoStop}{Gell-MannBrueckner-PR106}%
\bibitem{OnsagerMittagStephen-AnnPhysG473}%
  \BibitemOpen
  \bibfield{author}{%
  \bibinfo {author} {\bibfnamefont{L.}~\bibnamefont{Onsager}}, \bibinfo
  {author} {\bibfnamefont{L.}~\bibnamefont{Mittag}},\ and\ \bibinfo {author}
  {\bibfnamefont{M.~J.}\ \bibnamefont{Stephen}},\ }%
  \bibfield{journal}{%
  \Doi{10.1002/andp.19664730108}{\bibinfo {journal} {Annalen der Physik}}\ }%
  \textbf{\bibinfo {volume} {473}},\ \bibinfo {pages} {71} (\bibinfo {year}
  {1966})%
  \bibAnnoteFile{NoStop}{OnsagerMittagStephen-AnnPhysG473}%
\bibitem{IsiharaIoriatti-PRB22}%
  \BibitemOpen
  \bibfield{author}{%
  \bibinfo {author} {\bibfnamefont{A.}~\bibnamefont{Isihara}}\ and\ \bibinfo
  {author} {\bibfnamefont{L.}~\bibnamefont{Ioriatti}},\ }%
  \bibfield{journal}{%
  \Doi{10.1103/PhysRevB.22.214}{\bibinfo {journal} {Phys. Rev. B}}\ }%
  \textbf{\bibinfo {volume} {22}},\ \bibinfo {pages} {214} (\bibinfo {year}
  {1980})%
  \bibAnnoteFile{NoStop}{IsiharaIoriatti-PRB22}%
\bibitem{Glasser-JCAM10}%
  \BibitemOpen
  \bibfield{author}{%
  \bibinfo {author} {\bibfnamefont{M.~L.}\ \bibnamefont{Glasser}},\ }%
  \bibfield{journal}{%
  \Doi{10.1016/0377-0427(84)90041-4}{\bibinfo {journal} {J. Comput. Appl.
  Math.}}\ }%
  \textbf{\bibinfo {volume} {10}},\ \bibinfo {pages} {293} (\bibinfo {year}
  {1984})%
  \bibAnnoteFile{NoStop}{Glasser-JCAM10}%
\bibitem{ChakravortyDavidson-JPC100}%
  \BibitemOpen
  \bibfield{author}{%
  \bibinfo {author} {\bibfnamefont{S.~J.}\ \bibnamefont{Chakravorty}}\ and\
  \bibinfo {author} {\bibfnamefont{E.~R.}\ \bibnamefont{Davidson}},\ }%
  \bibfield{journal}{%
  \Doi{10.1021/jp952803s}{\bibinfo {journal} {J. Phys. Chem.}}\ }%
  \textbf{\bibinfo {volume} {100}},\ \bibinfo {pages} {6167} (\bibinfo {year}
  {1996})%
  \bibAnnoteFile{NoStop}{ChakravortyDavidson-JPC100}%
\bibitem{BrackBhaduri-1997}%
  \BibitemOpen
  \bibfield{author}{%
  \bibinfo {author} {\bibfnamefont{M.}~\bibnamefont{Brack}}\ and\ \bibinfo
  {author} {\bibfnamefont{R.~K.}\ \bibnamefont{Bhaduri}},\ }%
  \emph{\bibinfo {title} {{S}emiclassical {P}hysics}},\ Frontiers in Physics\
  (\bibinfo {publisher} {Addison-Wesley Publishing Company, Inc.},\ \bibinfo
  {year} {1997})%
  \bibAnnoteFile{NoStop}{BrackBhaduri-1997}%
\bibitem{BerryMount-RPP35}%
  \BibitemOpen
  \bibfield{author}{%
  \bibinfo {author} {\bibfnamefont{M.~V.}\ \bibnamefont{Berry}}\ and\ \bibinfo
  {author} {\bibfnamefont{K.~E.}\ \bibnamefont{Mount}},\ }%
  \bibfield{journal}{%
  \bibinfo {journal} {Rep. Prog. Phys.}\ }%
  \textbf{\bibinfo {volume} {35}},\ \bibinfo {pages} {315} (\bibinfo {year}
  {1972})%
  \bibAnnoteFile{NoStop}{BerryMount-RPP35}%
\bibitem{TaruchaEA-PRL77}%
  \BibitemOpen
  \bibfield{author}{%
  \bibinfo {author} {\bibfnamefont{S.}~\bibnamefont{Tarucha}}, \bibinfo
  {author} {\bibfnamefont{D.~G.}\ \bibnamefont{Austing}}, \bibinfo {author}
  {\bibfnamefont{T.}~\bibnamefont{Honda}}, \bibinfo {author}
  {\bibfnamefont{R.~J.}\ \bibnamefont{van~der Hage}},\ and\ \bibinfo {author}
  {\bibfnamefont{L.~P.}\ \bibnamefont{Kouwenhoven}},\ }%
  \bibfield{journal}{%
  \Doi{10.1103/PhysRevLett.77.3613}{\bibinfo {journal} {Phys. Rev. Lett.}}\ }%
  \textbf{\bibinfo {volume} {77}},\ \bibinfo {pages} {3613} (\bibinfo {year}
  {1996})%
  \bibAnnoteFile{NoStop}{TaruchaEA-PRL77}%
\end{thebibliography}
\end{document}